\documentclass[aps,prx,10pt,twocolumn,floatfix]{revtex4-1}

\usepackage{times}
\usepackage{graphicx}
\usepackage{subfigure}
\usepackage{dcolumn}
\usepackage{bm}
\usepackage{xcolor}
\usepackage{amsmath,amssymb}
\usepackage{verbatim}
\usepackage{mathtools}
\usepackage{epstopdf}
\usepackage{bbold}
\usepackage{changes}
\usepackage{soul}

\usepackage{hyperref}
\hypersetup{
colorlinks=true,
linkcolor=blue,           
citecolor=blue,           
filecolor=magenta,        
urlcolor=cyan
}

\renewcommand{\Re}{\mathrm{Re}\,}

\DeclarePairedDelimiter\ket{\lvert}{\rangle}
\DeclarePairedDelimiterX\braket[2]{\langle}{\rangle}{#1 \delimsize\vert #2}
\DeclarePairedDelimiterX\ketbra[2]{\lvert}{\rvert}{#1 \rangle\hspace{-.25em}\langle #2}

\begin{document}

\title{Topological framework for directional amplification in driven-dissipative cavity arrays}

\author{Clara C.~Wanjura}
\email{ccw45@cam.ac.uk}
\affiliation{Cavendish Laboratory, University of Cambridge, Cambridge CB3 0HE, United Kingdom}

\author{Matteo Brunelli}
\affiliation{Cavendish Laboratory, University of Cambridge, Cambridge CB3 0HE, United Kingdom}

\author{Andreas Nunnenkamp}
\affiliation{Cavendish Laboratory, University of Cambridge, Cambridge CB3 0HE, United Kingdom}

\date{\today}

\begin{abstract}
Directional amplification, in which signals are selectively amplified depending on their propagation direction, has attracted much attention as key resource for applications, including quantum information processing. Recently, several, physically very different, directional amplifiers have been proposed and realized in the lab. In this work, we present a unifying framework based on topology to understand non-reciprocity and directional amplification in driven-dissipative cavity arrays. Specifically, we unveil a one-to-one correspondence between a non-zero topological invariant defined on the spectrum of the dynamic matrix and regimes of directional amplification, in which the end-to-end gain grows exponentially with the number of cavities. We compute analytically the scattering matrix, the gain and reverse gain, showing their explicit dependence on the value of the topological invariant. Parameter regimes achieving directional amplification can be elegantly obtained from a topological `phase diagram', which provides a guiding principle for the design of both phase-preserving and phase-sensitive multimode directional amplifiers.
\end{abstract}


\maketitle

\section*{Introduction}

Controlling amplification and directionality of electromagnetic signals is one key resource for information processing. Amplification allows to compensate for attenuation losses and to read out signals while adding a minimal amount of noise. Directionality, also known as non-reciprocity, allows to select the direction of propagation while blocking signals in the reverse~\cite{Deak2012, Caloz2018}.
Non-reciprocity is of wide-ranging practical value; for instance, it simplifies the construction of photonic networks~\cite{Jalas2013, Ranzani2015, Metelmann2018}, enhances the information capacity in communication technology \cite{Miller2010, Verhagen2017}, and can be a resource for (quantum) sensing~\cite{Lau2018}.
Combining non-reciprocity and amplification, directional amplifiers allow for the detection of weak signals while protecting them against noise from the read-out electronics. For these reasons, these devices have become important components for promising quantum information platforms such as superconducting circuits~\cite{Abdo2011}.

In response to this demand, many proposals and realizations of non-reciprocal and amplifying devices have appeared in the recent literature.
Isolators and circulators based on magneto-optical effects 
have become the conventional choice, but they are bulky and require undesired magnetic fields to explicitly break time-reversal symmetry. Josephson junctions~\cite{Abdo2013DirAmp, Sliwa2015, Lecocq2017} have been investigated as an alternative. Other approaches include refractive index modulation~\cite{Yu2009, Lira2012}, interfering parametric processes~\cite{Kamal2011}, and optomechanics \cite{Manipatruni2009, Hafezi2012, Ruesink2016}.
An elegant solution is provided by reservoir engineering~\cite{Metelmann2014, Metelmann2015, Fang2017, Metelmann2017, Bernier2017, Peterson2017, Barzanjeh2017, Malz2018, Mercier2019}, where non-reciprocity is achieved by interfering coherent and dissipative processes~\cite{Metelmann2015, Metelmann2017}.
Based on this approach, several few-mode isolators and directional amplifiers have been proposed~\cite{Metelmann2014, Metelmann2015, Metelmann2017, Malz2018} and demonstrated~\cite{Fang2017, Bernier2017, Barzanjeh2017, Peterson2017, Mercier2019}.

\begin{figure}[h!]
\centering
\includegraphics[width=.48\textwidth]{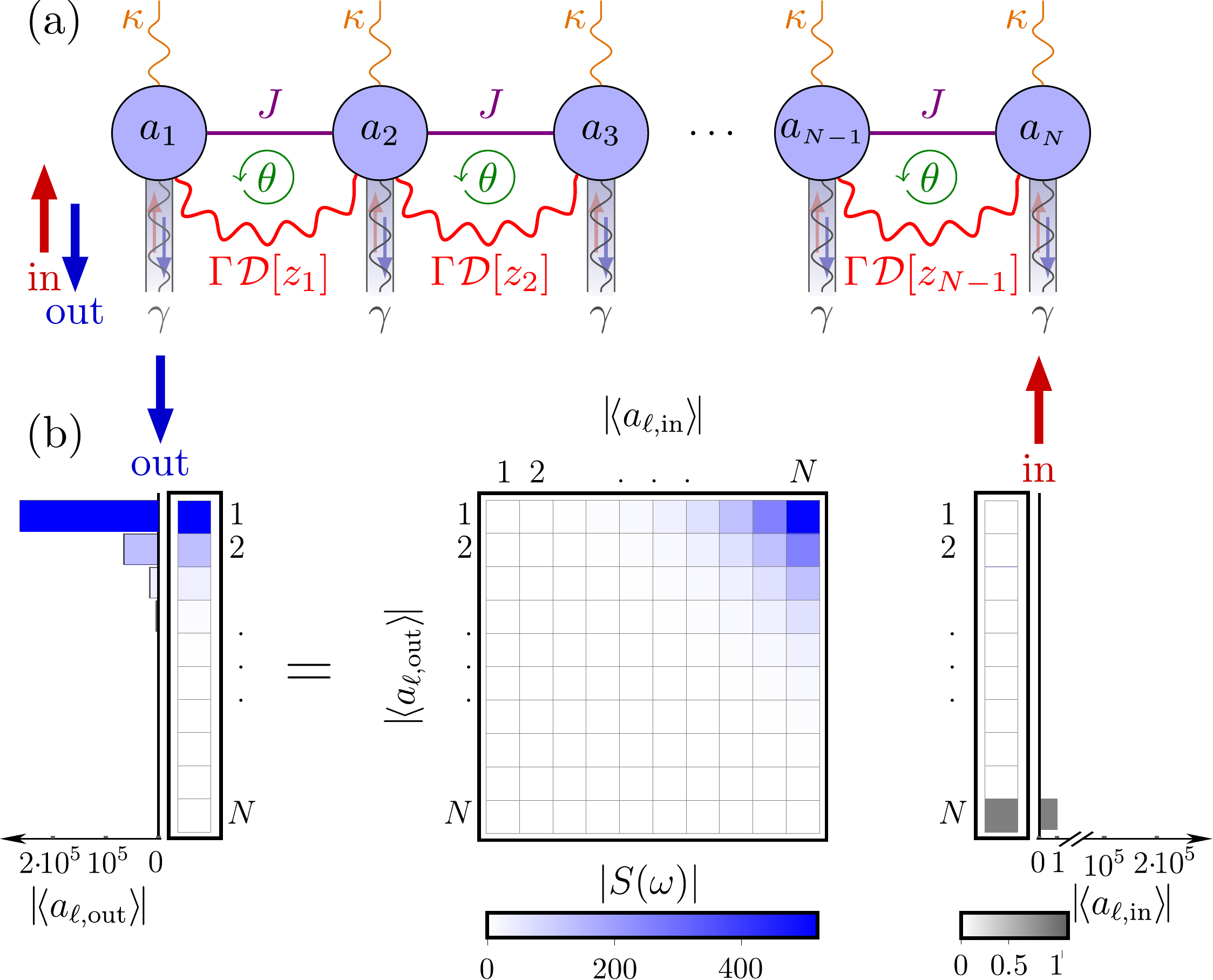}
\caption{
\textbf{Directional amplification in a driven-dissipative chain.}
(a) Driven-dissipative chain of $N$ bosonic cavity modes $a_j$ evolving according to Eq.~\eqref{eq:masterEq}. Neighboring modes are coupled both coherently with strength $J$, see Eq.~\eqref{eq:hamiltonian}, and dissipatively at rate $\Gamma$ through non-local dissipators $\mathcal{D}[z_j]$ with $z_j \equiv a_j + e^{-\mathrm{i}\theta} a_{j+1}$.
Each mode is coupled to a wave guide, which allows probing with a coherent input signal $\langle a_{\ell,\mathrm{in}}\rangle$ and introduces losses with rate $\gamma$. Incoherent pumping of photons at rate $\kappa$ enables an amplifying regime for which local dissipation overcomes non-local dissipation, see Eq.~\eqref{eq:dynEq}.
(b) Topologically non-trivial regimes of the chain correspond to non-reciprocal amplification of a coherent input signal.
In the topologically non-trivial regime $\nu \not=0$ (here $\nu=+1$), according to Eq.~\eqref{eq:winding}, an input at one end (right) exits amplified at the opposite end (left). This is quantified by the scattering matrix $S(\omega)$ (center) given by Eq.~\eqref{eq:scatMatIntroduction}.
From the structure of $S(\omega)$, we see that amplification is directional, i.e.~if input and output fields are exchanged, the transmission is strongly suppressed.
We have chosen $N=10$, $\mathcal{C}=2\Gamma/(\gamma+2\Gamma-\kappa)=2$, $\Lambda=4J/(\gamma+2\Gamma-\kappa)=2$ and $\theta=\frac{3\pi}{2}$.
}
\label{fig:introductory}
\end{figure}

On the other hand, chiral edge states of topological photonic systems~\cite{Ozawa2019} give rise to the directional transport of photons and phonons~\cite{Peano2015, Peano2016TopPhases}, which has been used to design traveling wave amplifiers~\cite{Peano2016} and topological lasers~\cite{St-Jean2017,Zhao2018,Harari2018, Bandres2018, Mittal2018}.
Transport phenomena in dissipative systems characterized by a topological winding number have been studied in \cite{Rudner2009, Kastoryano2019}.
A generalized winding number applied to a non-Hermitian system has previously appeared in the study of the Su-Schrieffer-Heeger (SSH) laser~\cite{Schomerus2013,Parto2018}.

In this paper, we unify the plethora of ad hoc proposals for directional amplifiers by uncovering an organizing principle underlying directional amplification in driven-dissipative cavity arrays: the non-trivial topology of the matrix governing the time evolution of the cavity modes. Based on this notion of topology, we develop a framework to understand directional amplification in multimode arrays and provide a recipe to design novel devices. The systems we consider are driven-dissipative cavity chains as the one depicted in Fig.~\ref{fig:introductory}~(a), featuring both coherent and dissipative couplings between modes. Non-trivial topology coincides with directional amplification and arises from the competition of local and non-local dissipative terms while the Hamiltonian describing the evolution of the closed system features a topologically trivial band structure.

We build our analysis on the scattering matrix illustrated in Fig.~\ref{fig:introductory}~(b). The scattering matrix characterizes the isolating properties as well as the amplification of a weak probe across the chain.
Next, we introduce a topological invariant, the winding number, see Fig.~\ref{fig:exampleTop}, which is defined on the spectrum of the dynamic matrix governing the evolution of the cavity amplitudes and enters directly in the scattering matrix.
We then employ the winding number to discuss the topological regimes of the driven-dissipative chain leading to the topological `phase diagram' for the scattering matrix, Fig.~\ref{fig:topPhaseDiag},
which at the same time defines the directionally amplifying parameter regimes.
We go on to rigorously prove the one-to-one correspondence between non-trivial topology and directional amplification leading to one of our main results: the analytic expression for the scattering matrix in non-trivial topological regimes, Eq.~\eqref{eq:mainResult}.
This result already holds for systems consisting of as few as two modes in the vicinity of the exceptional point (EP), where it is exact, and converges to the exact result exponentially fast within the whole topologically non-trivial regime.
From Eq.~\eqref{eq:mainResult} we find the exponential scaling of the amplifier gain with the chain length, Eq.~\eqref{eq:gainZpm}, while signals in the reverse direction are exponentially suppressed, Eq.~\eqref{eq:revGainZpm}.
Therefore, increasing the chain length enlarges the parameter range for which directional amplification occurs, from a fine-tuned point to the whole topologically non-trivial regime.
The generality of our results becomes clear in the last section of Results, in which we examine with our topological framework scaled-up versions of different models for phase preserving and phase sensitive amplifiers that have appeared in the literature~\cite{Metelmann2015, Metelmann2017, McDonald2018}. We demonstrate how we can predict the different amplifying regimes of these devices, compute gain and reverse gain, and obtain the scattering matrix from our topological framework by inspecting the winding number.
Directional amplification can be seen as a proxy of non-trivial topology, formally defined only in the thermodynamic limit, even in very small systems, which makes our work relevant for state-of-the art devices such as~\cite{Mercier2019}.

Our analysis serves as a general recipe for designing multimode amplifiers that can be integrated in scalable platforms, such as superconducting circuits~\cite{Bergeal2010, Abdo2013DirAmp}, optomechanical systems~\cite{Aspelmeyer2014}, and topolectric circuits~\cite{Lee2018, Kotwal2019}.
Finally, our work also has direct relevance for the study of the topology of non-Hermitian Hamiltonians~\cite{MartinezAlvarez2018,Porras2018}, for which similar topological invariants have been proposed~\cite{Gong2018, Ghatak2019}, leading to the recent classification in terms of 38 symmetry classes~\cite{Kawabata2018}.
In this context, our work provides a direct way to detect topological features, e.g. extract the value of the topological invariant, which has previously been challenging.

\section*{Results}
\subsection*{Directional amplification in a driven-dissipative chain}
\label{sec:chain}

Let us start by introducing the system that will guide us through the general discussion and illustrate our results.
We consider a driven-dissipative chain of $N$ identical cavity modes $a_j$ as depicted in Fig.~\ref{fig:introductory}~(a).
Its coherent evolution in a frame rotating with respect to the cavity frequency is governed by the Hamiltonian ($\hbar=1$)
\begin{align}
\mathcal{H} = \sum_j (J a_j^\dagger a_{j+1}+ J^*a_j a_{j+1}^\dagger),
\label{eq:hamiltonian}
\end{align}
which describes photons hopping with uniform amplitude $J$ along the chain.
The dissipation consists of both local and non-local contributions and is described by the master equation
\begin{align}
   \dot \rho = -\mathrm{i} [\mathcal{H},\rho]
      + \sum_j \left(
         \Gamma \mathcal{D}[z_j]\rho + \gamma \mathcal{D}[a_j]\rho + \kappa \mathcal{D}[a_j^\dagger]\rho
         \right) \label{eq:masterEq}
\end{align}
for the system density matrix $\rho$.  The first dissipator $\mathcal{D}[z_j]\rho=z_j\rho z_j^\dagger-\frac{1}{2}\{z_j^\dagger z_j,\rho\}$ with $z_j\equiv a_j+e^{-\mathrm{i}\theta}a_{j+1}$ couples dissipatively neighboring cavities with rate $\Gamma$~\cite{Porras2018, Metelmann2015}, the second describes photon decay into the wave guide with rate $\gamma$, while the last is an incoherent pump at rate $\kappa$. This last term can be implemented with the help of a parametrically coupled auxiliary mode which is subsequently adiabatically eliminated from the equations of motion.
The phase $\theta$ can for instance be obtained in a driven optomechanical setup~\cite{Aspelmeyer2014,Bernier2017,Malz2018},
in which the mechanical mode is adiabatically eliminated giving rise to the non-local dissipator.
The controllable phase of the pumps is imprinted onto the amplitude of the coherent state inside the cavities and therefore transferred to the optomechanical coupling constant. This gives rise to the phase $\theta$.

Our main interest will be in the fields entering $\langle a_{j,\mathrm{in}}(t)\rangle$ and exiting $\langle a_{j,\mathrm{out}}(t)\rangle$ the cavities through the wave guides, which are connected via the input-output boundary conditions $\langle a_{j,\mathrm{out}}\rangle=\langle a_{j,\mathrm{in}}\rangle + \sqrt{\gamma}\langle a_j\rangle$~\cite{Gardiner1985, Clerk2010}.

Following the standard procedures, we obtain the following equations of motion for the cavity amplitudes $\langle a_j\rangle$
\begin{align}
\langle\dot a_j\rangle = &
\frac{\kappa-\gamma-2\Gamma}{2}  \langle a_j\rangle - \sqrt{\gamma} \langle a_{j,\mathrm{in}}\rangle \notag \\
& - \left(\mathrm{i} J + \frac{e^{-\mathrm{i}\theta}\Gamma}{2}\right) \langle a_{j+1}\rangle - \left(\mathrm{i} J+\frac{e^{\mathrm{i}\theta}\Gamma}{2}\right)  \langle a_{j-1}\rangle \notag \\
\equiv & \sum_j H_{j,\ell} \langle a_\ell\rangle - \sqrt{\gamma} \langle a_{j,\mathrm{in}}(t)\rangle.
\label{eq:dynEq}
\end{align}

In these equations (\refeq{eq:dynEq}), we have chosen $J$ real, which is always possible due to gauge freedom~\cite{Metelmann2015}. The input $\langle a_{j,\mathrm{in}}(t)\rangle$ enters as a coherent drive in the frame rotating with the cavity frequency. Note that the non-local dissipator contributes both to the coupling terms and to the local decay rate. The phase $\theta$ is crucial for the non-reciprocity of the chain: since coherent and dissipative couplings between neighboring modes form a closed path, these processes can interfere constructively or destructively depending on the phase~$\theta$. For example, setting $\mathrm{i} J=-e^{\mathrm{i}\theta}\Gamma/2$, i.e.~$\theta=\frac{3\pi}{2}$, in Eqs.~\eqref{eq:dynEq}, each cavity $j$ in Fig.~\ref{fig:introductory}~(a) only couples to its right-hand side neighbor $(j+1)$, but not to the cavity $(j-1)$ on its left. This leads to the complete cancellation of the transmission from left to right~\cite{Metelmann2015, Metelmann2017} and corresponds to standard cascaded quantum systems theory~\cite{Carmichael1993, Gardiner1993}.
These are also the exceptional points of the system as we show in Methods.

As we can see from the last line of Eqs.~\eqref{eq:dynEq}, the evolution equations can be conveniently expressed as matrix-vector product with $H$ the dynamic matrix. $H$ plays an important role in characterizing the transmitting and amplifying properties of the system. This is because it determines the scattering matrix $S(\omega)$, which linearly links the input $\langle a_{j,\mathrm{in}}(\omega)\rangle$ to the output fields $\langle a_{j,\mathrm{out}}(\omega)\rangle$ in frequency space
\begin{align}
{\bf a_\mathrm{out}} & = [\mathbb{1}+\gamma(\mathrm{i}\omega\mathbb{1}+H)^{-1}] {\bf a_\mathrm{in}} \equiv S(\omega) {\bf a_\mathrm{in}},
\label{eq:scatMatIntroduction}
\end{align}
where we set ${\bf a_\mathrm{in/out}}\equiv(\langle a_{1,\mathrm{in/out}}\rangle,\dots,\langle a_{N,\mathrm{in/out}}\rangle)^\mathrm{T}$.

Fig.~\ref{fig:introductory}~(b) illustrates the role of the scattering matrix for the driven-dissipative chain. As we can see, the chain acts as a directional amplifier in the case shown: the dominant top right corner of $S(\omega)$ relates a weak input signal at the $N$th cavity to a strongly amplified output at the first cavity, while transmission in the opposite direction is suppressed. Formally, non-reciprocity between modes $j$ and $\ell$ corresponds to the condition $\lvert S_{j,\ell}\rvert \neq \lvert S_{\ell,j}\rvert$ and practically useful amplification to $\lvert S_{j,\ell}\rvert\gg 1$.

Indeed, one of the key quantities used to characterize amplifiers is the gain $\mathcal{G}$~\cite{Clerk2010}, which we define as the scattering matrix element with the largest absolute value. For the driven-dissipative chain, the gain relates the input at the first (last) to the output at the last (first) cavity as follows
\begin{align}
\mathcal{G}(\omega) \equiv
\begin{cases} \lvert S_{N,1}(\omega)\rvert^2: & \theta\in(0,\pi) \\ \lvert S_{1,N}(\omega)\rvert^2: & \theta\in(\pi,2\pi).\end{cases} \label{eq:gain}
\intertext{Conversely, the reverse gain pertains to the transmission in the opposite propagation direction}
\bar{\mathcal{G}}(\omega)\equiv\begin{cases} \lvert S_{1,N}(\omega)\rvert^2: & \theta\in(0,\pi) \\ \lvert S_{N,1}(\omega)\rvert^2: & \theta\in(\pi,2\pi).\end{cases} \label{eq:revGain}
\end{align}
An efficient directional amplifier obeys $\mathcal{G}\gg1$ and $\bar{\mathcal{G}}\ll1$.

For convenience, we introduce
\begin{align}
M(\omega)\equiv\mathrm{i}\omega\mathbb{1}+H \label{eq:M}
\end{align}
with $M(0)=H$ and dub it dynamic matrix at frequency $\omega$.
We also define its inverse as the susceptibility matrix
\begin{align}
\chi(\omega)\equiv(\mathrm{i}\omega\mathbb{1}+H)^{-1}, \label{eq:chi}
\end{align}
which is related to the scattering matrix through
\begin{align}
S(\omega) & = \mathbb{1}+\gamma \chi(\omega). \label{eq:scatMat}
\end{align}
It is clear that $M(\omega)$ determines the properties of $S(\omega)$ and we use it to define a topological invariant.

\subsection*{The winding number}
\label{sec:winding}

In this section, we introduce a topological invariant akin to the winding number of the canonical SSH model~\cite{Asboth2016}, but defined on the complex spectrum of the dynamic matrix (in reciprocal space). The same topological invariant was recently studied by Gong et al.~\cite{Gong2018} for non-Hermitian Hamiltonians.

In general, the dynamic matrix of a translational invariant 1D system, such as our driven-dissipative chain, has the form $M_{j,j+\ell}\equiv\mu_\ell$ for all $j$.
Our strategy is to employ periodic boundary conditions (PBC) to probe the bulk properties and to define a meaningful topological invariant --- the winding number.
We will see that the system is extremely sensitive to changes of the boundary conditions. Indeed, moving to open boundary conditions (OBC) leads to the directional amplification we want to characterize.

Under PBC, $M_\mathrm{pbc}$ is diagonal in the plane wave basis ${\ket{k}=\frac{1}{\sqrt{N}}\sum_j e^{\mathrm{i} k j}\ket{j}}$ with $k=2\pi r/N$, $r=0,1,\dots,N-1$
\begin{align}
M_\mathrm{pbc}
& = \sum_{\ell}\mu_\ell\sum_j \ketbra{j}{(j+\ell)\bmod N} \notag \\
& = \sum_k\sum_\ell \mu_\ell e^{\mathrm{i} k \ell} \ketbra{k}{k}
\equiv \sum_k h(k) \ketbra{k}{k}, \label{eq:Mk}
\end{align}
with the generating function ${h(k)\equiv \sum_\ell \mu_\ell e^{\mathrm{i} k \ell}}$.
Equivalently, $h(k)$ generates the entries ${\mu_\ell=\frac{1}{2\pi}\int_0^{2\pi}\mathrm{d}k\, h(k) e^{-\mathrm{i} k \ell}}$ of $M$.
We have adopted a Dirac notation for referring to the (cavity) site basis $\{\ket{j}\}$ and plane wave basis $\{\ket{k}\}$, respectively.

\begin{figure}[t]
\centering
\includegraphics[width=.48\textwidth]{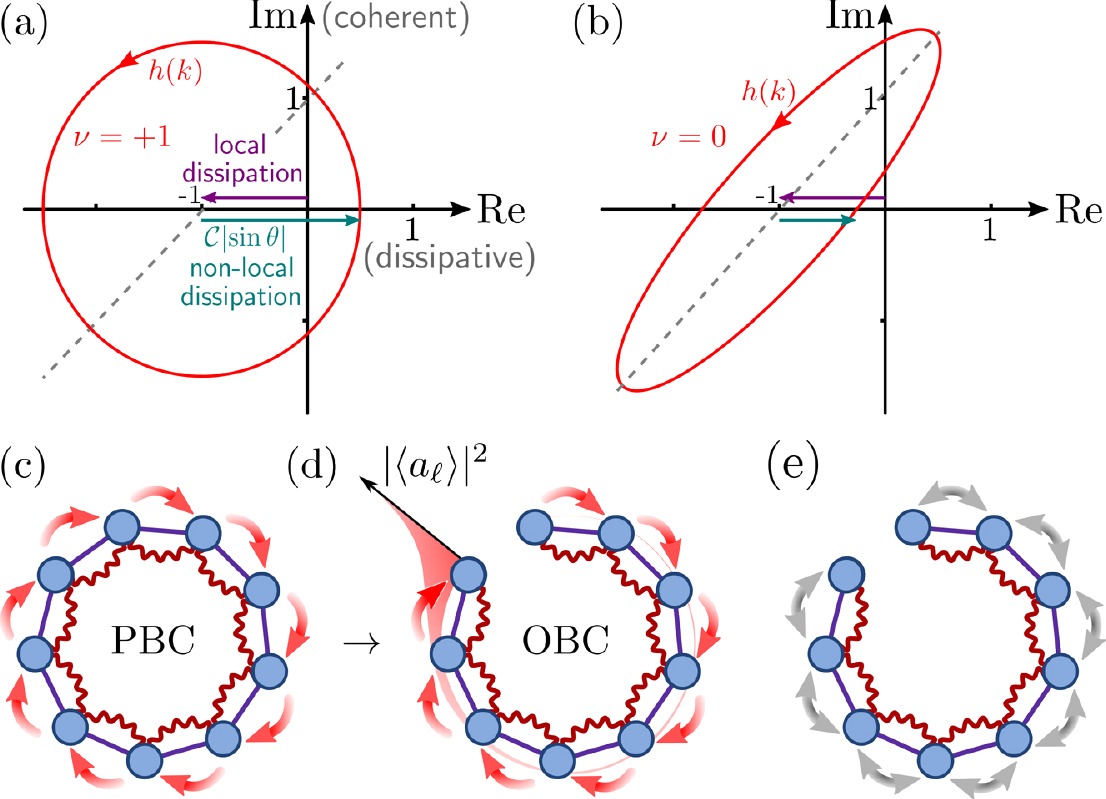}
\caption{
\textbf{Topological invariant for the dynamic matrix, and intuitive explanation for the gain.}
Under PBC, the eigenvalues of the dynamic matrix $M(\omega)$, Eq.~\eqref{eq:M}, describe a closed curve $h(k)$ (red) in the complex plane --- the generating function~\eqref{eq:Mk}. This allows us to define the winding number $\nu$ of Eq.~\eqref{eq:winding} counting the revolutions of $h(k)$ around the origin.
(a)~On resonance, $\omega=0$, the non-local dissipation has to surpass the local dissipation to yield a non-trivial winding number, see Eq.~\eqref{eq:genFuncChain}. Otherwise, (b) $\nu$ is trivial. This competition between local and non-local contributions in the generating function is indicated by the purple and blue arrows. For $\theta=0$ or $\pi$, $h(k)$ degenerates into a line, which is shown for $\theta=0$ in (a) and (b) as dashed lines. When $\theta=\pi$ the slope  changes sign.
In (a) $\theta=\frac{\pi}{2}$ and in (b) $\theta=0.5$.
(c)~Under PBC with $\nu\neq0$, excitations travel directionally around the ring and gain energy at each revolution causing instability. (d)~Removing one link (OBC) leads to the accumulation of excitations at one end, which determines the end-to-end gain.
(e)~For reciprocal dynamics, removing the link only induces local changes and no gain.}
\label{fig:exampleTop}
\end{figure}

$h(k)$ can be regarded as an energy band in the 1D Brillouin zone; only that now, $h(k)$  takes complex values since $M\neq M^\dagger$.
As $h(k)$ is periodic in $k$ with period $2\pi$, it describes a closed curve in the complex plane, cf. Fig.~\ref{fig:exampleTop}. This enables us to define a winding number from the argument principle~\cite{Gong2018}
\begin{align}
\nu \equiv \frac{1}{2\pi\mathrm{i}} \int_0^{2\pi} \mathrm{d}k \, \frac{h'(k)}{h(k)} = \frac{1}{2\pi\mathrm{i}} \oint_{\lvert z\rvert=1} \mathrm{d}z \, \frac{\frac{\partial}{\partial z} h(z)}{h(z)}, \label{eq:winding}
\end{align}
where we have introduced $z=e^{\mathrm{i} k}$ in the last step.
The winding number is an integer counting the number of times $h$ wraps around the origin as $k$ changes from $0$ to $2\pi$.
While Gong et al.~[43] define the winding number w.r.t.~an arbitrary base point, we choose the origin as special point for the physically relevant scattering matrix: as we will see later from~Eq.~(18), it is the pole of the scattering matrix under PBC.

\begin{figure*}[t]
\centering
\includegraphics[width=.9\textwidth]{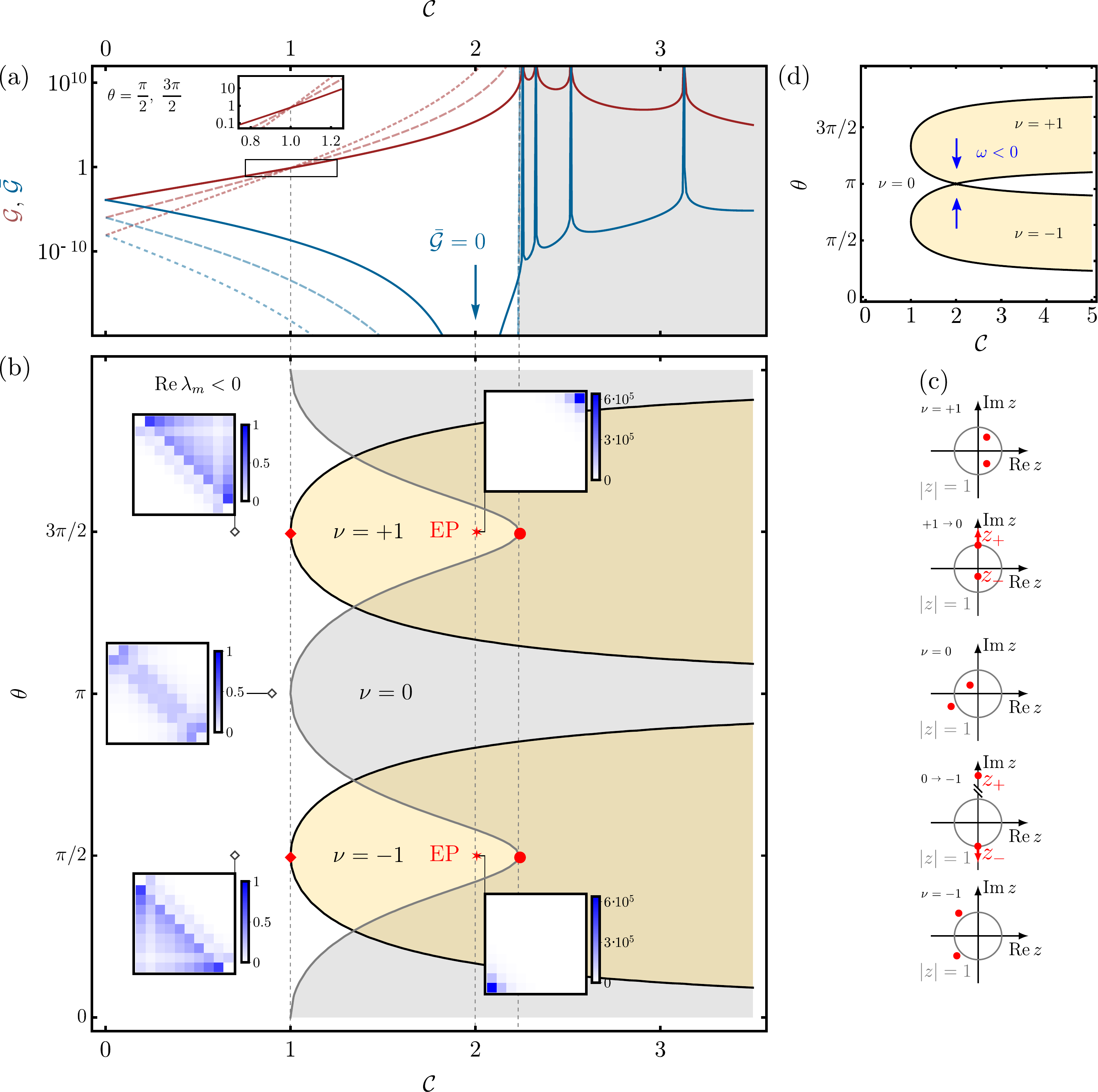}
\caption{
\textbf{Topological `phase diagram' of the scattering matrix.}
(a) Gain $\mathcal{G}(0)$ and reverse gain $\bar{\mathcal{G}}(0)$, see Eqs.~\eqref{eq:gain} and~\eqref{eq:revGain}, respectively, for $N=10$ (solid line), $N=15$ (dashed), $N=20$ (dotted), all for $\theta=\frac{\pi}{2},\frac{3\pi}{2}$, and (b) general topological `phase diagram' on resonance, $\omega=0$, with distinct winding numbers according to Eq.~\eqref{eq:winding}.
We can associate a scattering matrix $S(0)$ with each point in the diagram and we show some $\lvert S(0)\rvert^2$ as inset with $\Lambda=2$ and $\gamma=2\Gamma-\kappa$ in Eq.~\eqref{eq:scatMat} to obtain impedance matching at the exceptional point (EP). Note in particular the color scales of the scattering matrices revealing the amplification, and the asymmetry of the matrix signifying non-reciprocity.
Condition~\eqref{eq:topCondNoDet} yields the orange lobes in (b) and corresponds to winding numbers $\nu=\pm1$, whereas the rest is the trivial regime $\nu=0$. Directional amplification, i.e.~$\mathcal{G}>1$, sets in as we move into a topologically non-trivial regime. For the parameters shown in (a) this occurs at $\mathcal{C}=1$. In this regime, the gain grows exponentially with $N$.
At the EP the transmission in the reverse direction is completely suppressed, i.e.~$\bar{\mathcal{G}}=0$, and the upper (lower) triangle of $S(0)$ is exactly zero.
The system becomes unstable (gray overlay), when $\max_m\mathrm{Re}\,\lambda_m>0$, in which $\lambda_m$ is the $m$th eigenvalue of the dynamic matrix $M_\mathrm{obc}(0)$, see Eqs.~\eqref{eq:dynEq} and~\eqref{eq:eigenvals}. $\mathrm{Re}\,\lambda_m=0$ coincides with the onset of the  parametric instability and can be seen as divergence in the gain in (a).
Non-reciprocity also occurs outside of non-trivial topological regimes and is governed by the phase $\theta$. Complete directionality is achieved at $\theta=\frac{\pi}{2}$ for $\nu=-1$ from left to right ($\frac{3\pi}{2}$ for $\nu=+1$ from right to left).
While the gain only depends weakly on $\Lambda$, larger (smaller) $\Lambda$ shifts the location of the EP to the right (left) and extends (shrinks) the stable regime.
(c) The number of zeros inside the unit circle determines the winding number. On the boundary between trivial and non-trivial regimes, one of the zeros lies on the unit circle and hence $\mathcal{G}=\mathcal{O}(1)$ independent of $N$, see (a) at $\mathcal{C}=1$.
(d) Off-resonance, $\omega\neq0$ shifts the two lobes corresponding to non-trivial topological regimes $\nu=\pm1$ towards each other. Where they overlap, we obtain a trivial regime.
}
\label{fig:topPhaseDiag}
\end{figure*}

In the following, we focus on nearest-neighbor interactions between cavity modes. Mathematically, this translates into generating functions of the form
\begin{align}
h(k) & = \mu_0 + \mu_1 e^{\mathrm{i} k} + \mu_{-1} e^{-\mathrm{i} k}
       = \mu_0 + \mu_1 z + \mu_{-1} \frac{1}{z}
\label{eq:genFunction}
\end{align}
permitting only $\nu=0,\pm1$.
In the Hermitian case, a Hamiltonian without any additional symmetries would be topologically trivial. However, in the case of non-Hermitian operators, one complex band is enough to obtain non-trivial values of a topological invariant~\cite{Gong2018}.
Note that Eq.~(\ref{eq:winding}) connects the winding number to the number of zeros of $h(z)$ encloses within the unit circle. 
For nearest-neighbor interactions, the zeros  are given by
\begin{align}
z_\pm\equiv \frac{-\mu_0\pm\sqrt{\mu_0^2-4\mu_1 \mu_{-1}}}{2\mu_1}. \label{eq:zpm}
\end{align}
The values of $\nu$ defining different topological regimes correspond to having two zeros within the unit circle ($\nu=+1$), one zero ($\nu=0$), or none ($\nu=-1$), see Fig.~\ref{fig:topPhaseDiag}~(c). Due to the form of Eq.~\eqref{eq:zpm} it is clear that non-trivial topological regimes are always linked to the competition of local, i.e.~$\mu_0$, and non-local terms, $\mu_1\mu_{-1}$.

\subsection*{Topological regimes of the driven-dissipative chain}
\label{sec:topRegimes}

We first consider the resonant response $\omega=0$, i.e.~when the probe frequency equals the cavity frequency. It is convenient to rescale all parameters by the on-site decay rate $(\gamma+2\Gamma-\kappa)/2$ and we introduce a rescaled hopping constant $\Lambda\equiv4J/(\gamma+2\Gamma-\kappa)$ and a cooperativity $\mathcal{C}\equiv2\Gamma/(\gamma+2\Gamma-\kappa)$ defined analogous to~\cite{Metelmann2015}. $\mathcal{C}$ is the ratio between the non-local dissipative contributions $\Gamma$ in Eqs.~\eqref{eq:dynEq} and the overall on-site decay rate $(\gamma+2\Gamma-\kappa)/2$. We refer to these two terms as non-local and local dissipation, respectively.

With these definitions, the generating function~\eqref{eq:genFunction} obtained from Eqs.~\eqref{eq:dynEq} becomes
\begin{align}
h(k) \propto -1 - \mathcal{C}\cos(k+\theta) - \mathrm{i}\Lambda\cos k.
\label{eq:genFuncChain}
\end{align}
We have dropped the proportionality factor $(\gamma+2\Gamma-\kappa)/2$ since the winding number is unchanged by the multiplication of the generating function with a non-zero constant.
Fig.~\ref{fig:exampleTop}~(a) and (b) illustrate $h(k)$ in the complex plane in topologically non-trivial and trivial regimes, respectively. Eq.~\eqref{eq:genFuncChain} shows that the imaginary part of $h(k)$ pertains to the coherent evolution, while the real part encodes the dissipation. Therefore, the winding number \eqref{eq:winding} is only well-defined in the presence of dissipation.
The imaginary part of $h(k)$ in Eq.~\eqref{eq:genFuncChain} takes both positive and negative values, so any non-vanishing $\Lambda$ can lead to $\nu\neq0$.
However, the real part in Eq.~\eqref{eq:genFuncChain} contains a constant shift ($-1$), which is due to local dissipation. This implies that the oscillating contribution $\mathcal{C}\cos(k+\theta)$ from the non-local dissipative interaction needs to exceed this local contribution to include the origin within $h(k)$, cf. Fig.~\ref{fig:exampleTop}.
A non-trivial winding number therefore always requires
\begin{align}
\mathcal{C}^2\sin^2\theta>1 \label{eq:topCondNoDet}
\end{align}
for $\nu\neq0$.
This yields the `phase diagram' Fig.~\ref{fig:topPhaseDiag}~(b) with the two orange lobes $\nu=\pm1$.
We note that $\nu\neq0$ is inaccessible for reciprocal dynamics ($\theta=0,\pi$). In this case, $h(k)$ degenerates into a line in the complex plane and $\nu=0$, unless it crosses the origin, in which case the winding number becomes undefined.

Entering the non-trivial topological regime is only possible with the help of the incoherent pump $\mathcal{D}[a_j^\dagger]$ of rate $\kappa$ in Eq.~\eqref{eq:masterEq} featuring as local anti-damping in Eqs.~\eqref{eq:dynEq}.
Condition~(\ref{eq:topCondNoDet}) implies that we require at least $1<\mathcal{C}=1/(1+\frac{\gamma-\kappa}{2\Gamma})$, which is equivalent to $\kappa>\gamma$. Hence, the modes $a_j$ have to be coupled to a bath which is out of equilibrium to obtain $\nu\neq0$.

The system response is captured by the scattering matrix $S(0)$, for which we show some representative examples under OBC within different regimes as insets in Fig.~\ref{fig:topPhaseDiag}~(b). Indeed, we can associate a scattering matrix with each point in the `phase diagram' and obtain qualitatively the same behavior within one topological regime.

Fig.~\ref{fig:topPhaseDiag}~(a) shows gain and reverse gain under OBC for $\theta=\frac{\pi}{2},\frac{3\pi}{2}$. End-to-end amplification sets in for $\mathcal{C}>1$ as we enter the topologically non-trivial regime, while transmission in the reverse direction is strongly suppressed. The sign of $\nu$ sets the propagation direction: $\nu=+1$ ($\nu=-1$) leads to amplification from right (left) to left (right). In regimes with $\nu=0$, the gain dominates over the reverse gain, but no amplification takes place.
This is a clear indication that non-trivial winding numbers coincide with directional amplification.
Note that within topologically non-trivial regimes the gain grows exponentially with $N$, $\mathcal{G}_{\nu=\pm1}\propto \lvert z_\mp\rvert^{-2\nu N}$ (for $N\gg1$) --- a result we will derive in the next section.

At the transition from the trivial to the non-trivial regime, the corresponding $z_\pm$ is located on the unit circle, see Fig.~\ref{fig:topPhaseDiag}~(c). Therefore, the gain is asymptotically independent of $N$ and $\mathcal{O}(1)$, see Fig.~\ref{fig:topPhaseDiag}~(a).
Within regimes $\nu\neq0$, the gain increases with $\mathcal{C}$ while the reverse gain decreases until we reach the EP $\mathcal{C}=\Lambda$, and $\theta=\frac{\pi}{2}$ or $\frac{3\pi}{2}$, at which $\bar{\mathcal{G}}=0$.
Note that $\Lambda$ sets the position of the EP on the lines $\theta=\frac{\pi}{2}$ and $\theta=\frac{3\pi}{2}$. For $\Lambda>1$ it is located within the topologically non-trivial regime, which is advantageous for a directional amplifier.

Our driven-dissipative chain not only cancels the signal in the reverse direction, it also ensures that any field entering the output cavity is not back-reflected and mixed-in with the output signal since we can choose $\gamma$ in Eq.~\eqref{eq:scatMat} such that $S_{1,1}=S_{N,N}=0$ (impedance matching) whenever $\theta=\frac{\pi}{2}$ or $\frac{3\pi}{2}$ in the stable regime, see insets in Fig.~\ref{fig:topPhaseDiag}~(b).
At the EP, the condition for impedance matching can be found analytically as $\gamma=2\Gamma-\kappa$.
This is a significant advantage over other proposals for directional amplifiers which do not necessarily have this property~\cite{Metelmann2015, Metelmann2017, Malz2018}. Among other things, it means that the amplifier is phase preserving even if signals are scattered back from other devices behind the amplifier.

The gain continues to increase with larger $\mathcal{C}$ beyond the exceptional point until we reach the parametric instability at which one eigenvalue of $M_\mathrm{obc}$ is zero.
We have an analytic expression for the eigenvalues under OBC available~\cite{Willms2008}, which we provide in Methods and use to plot the unstable regime in Fig.~\ref{fig:topPhaseDiag}~(a) and (b); all other regimes are stable.

Crucially, a longer chain also leads to the suppression of the reverse gain. Indeed, the reverse gain scales inversely with respect to $\mathcal{G}$, i.e.~$\bar{\mathcal{G}}_{\nu=\pm1}\propto \lvert z_\pm \rvert^{2\nu N}$, and $\bar{\mathcal{G}}$ vanishes at the EP, see Eq.~\eqref{eq:revGainZpm} and Fig.~\ref{fig:topPhaseDiag}~(a). This improves the isolation considerably, and in the thermodynamic limit, $N\to\infty$, extends the parameter regime over which we obtain completely directional amplification from the fine-tuned EP to the entire non-trivial topological regime.

Directional amplification is induced by the transition from PBC to OBC, which can intuitively be understood as follows:
For PBC and $\nu\neq0$, excitations travel around the ring in a given direction gaining energy, see Fig.~\ref{fig:exampleTop}~(c). In this case, the dynamics are unstable, since the eigenvalues $h(k)$ need to have both positive and negative real part to encircle the origin, see Fig.~\ref{fig:exampleTop}~(a).
Removing one link (OBC) can lead to stable dynamics and to the accumulation of excitations at one end of the chain, which translates into amplified steady state cavity amplitudes $\lvert\langle a_\ell\rangle\rvert^2$, see Fig.~\ref{fig:exampleTop}~(d).
For reciprocal dynamics, OBC only lead to local changes and no directional amplification, see Fig.~\ref{fig:exampleTop}~(e).

On resonance, the existence of non-trivial topological regimes is independent of the coherent coupling $\Lambda\neq0$. This changes, for the non-resonant response $\omega\neq0$. Rescaling also $\omega$ accordingly, $\tilde\omega\equiv2\omega/(\gamma+2\Gamma-\kappa)$, we obtain
\begin{align}
h(k) \propto & - 1 + \mathrm{i} \tilde\omega - \mathcal{C}\cos(k+\theta) - \mathrm{i}\Lambda\cos k.
\end{align}
Local and non-local contributions in both real and imaginary parts compete to yield a non-zero winding number. The condition for non-trivial topology reads
\begin{align}
\left(\frac{1}{\mathcal{C}\sin\theta}-\frac{\tilde\omega}{\Lambda\tan\theta}\right)^2+\frac{\tilde\omega^2}{\Lambda^2}<1.
\end{align}
This amounts to shifting the two lobes $\nu=\pm1$ against each other whereby the overlapping region becomes trivial, see Fig.~\ref{fig:topPhaseDiag}~(d).

\subsection*{One-to-one correspondence of non-trivial topology and directional amplification}
\label{sec:TopAmpCorrespondence}

We now rigorously prove the existence of a one-to-one correspondence between non-trivial values of the winding number and directional amplification for generic 1D systems with nearest-neighbor interactions that give rise to a dynamic matrix of Toeplitz form with uniform coupling constants.
To establish the correspondence, we study the susceptibility $\chi(\omega)=M^{-1}(\omega)$, first under PBC and then under OBC. Within non-trivial topological regimes, the corrections that arise from moving to OBC, lead to directional amplification by several orders of magnitudes. While we focus on nearest-neighbor couplings and generating functions of the form~\eqref{eq:genFunction}, our technique can also be employed beyond nearest-neighbor interactions.

Under PBC, calculating $\chi_\mathrm{pbc}$ is straightforward. For clarity, we omit the argument $\omega$ in what follows. Since we are ultimately interested in the scattering matrix, we express $\chi_\mathrm{pbc}=M_\mathrm{pbc}^{-1}$ in the site basis
\begin{align}
   \chi_\mathrm{pbc}
   & = \sum_k \frac{1}{h(k)} \ketbra{k}{k} = \sum_{j,\ell} \frac{1}{N} \sum_k \frac{e^{\mathrm{i} k (j-\ell)}}{h(k)} \ketbra{j}{\ell}. \label{eq:MpbcInv}
\end{align}
We see now, why the origin is a special point in the complex plane: it constitutes the pole of the scattering matrix.

\begin{figure}[t]
\centering
\includegraphics[width=.48\textwidth]{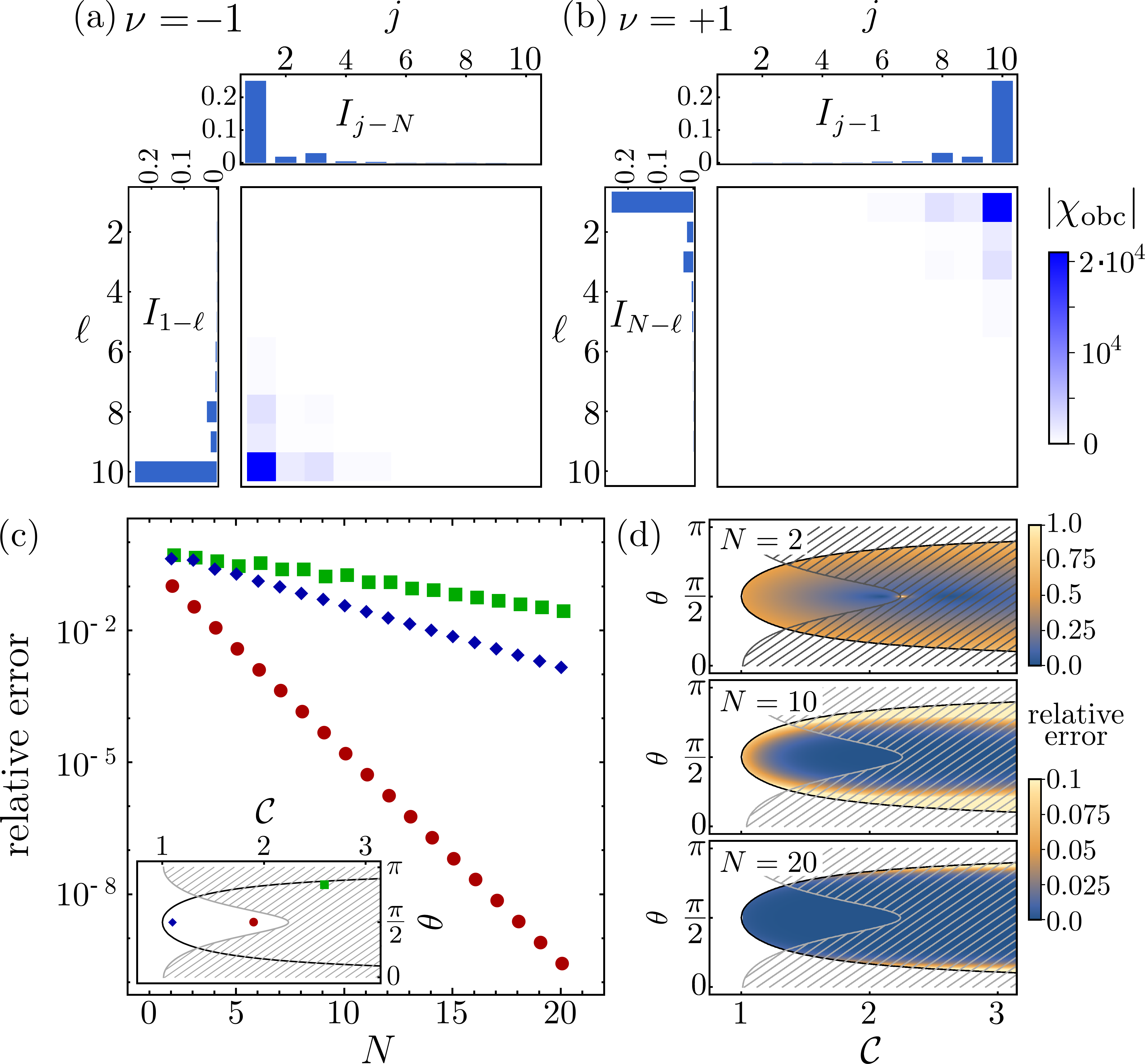}
\caption{
\textbf{Non-reciprocity and amplification in the susceptibility, and convergence of~Eq.~\eqref{eq:mainResult}.}
(a) $\nu=+1$ and (b) $\nu=-1$. The susceptibility $\chi_\mathrm{obc}$ is dominated by the middle sum in~(\ref{eq:matrixInverseTop}) and (\ref{eq:matrixInverseTop2}): products of the functions $I_n$ lead to the asymmetry of the scattering matrix with one dominant matrix element indicating non-reciprocity, whereas the amplification is determined by the pre-factor $1/\varepsilon_{\nu(1-N)}$, Eq.~\eqref{eq:epsilonExact}, and grows exponentially with the number of cavities~$N$. (a)~$\mu_0=0.3$, $\mu_1=0.5$, $\mu_{-1}=4$ and (b) $\mu_0=0.3$, $\mu_1=4$, $\mu_{-1}=0.5$.
(c)~Relative error of $\lvert(\chi_\mathrm{obc})_{N,1}\rvert$, Eq.~\eqref{eq:matrixInverseTop2}, for different points in the topological `phase diagram'. All show exponential convergence with the fastest close to the EP, and a slower rate in the dynamically unstable regime (hatched region) and at the boundary.
(d)~Relative error of $\lvert(\chi_\mathrm{obc})_{N,1}\rvert$ in the regime $\nu=-1$.
For $N=10$ and $N=20$ the color scale is cut at $0.1$.
The case $\nu=+1$ is analogous.}
\label{fig:MobcInverse}
\end{figure}

Rewriting the sum over $k$, we make the connection to the zeros of the generating function and hence $\nu$. For this purpose, we expand $z^{j-\ell}/h(z)$ into a Laurent series around $z=0$
\begin{align*}
\frac{z^{j-\ell}}{h(z)}
& = \frac{1}{2\pi\mathrm{i}}\sum_{n=-\infty}^\infty z^n \oint_{\lvert \tilde z\rvert=1} \mathrm{d} \tilde z \, \frac{\tilde z^{(j-\ell)-n-1}}{h(\tilde z)}.
\end{align*}
Inserting this expression into Eq.~(\ref{eq:MpbcInv}) allows us to evaluate the sum over $k$. Since $z=e^{\mathrm{i} k}$ and $k=2\pi r/N$ takes discrete values, we can write
\begin{align*}
   \chi_\mathrm{pbc}
      & = \sum_{j,\ell} 
          \sum_{n=-\infty}^\infty \sum_{r=1}^N \frac{e^{\mathrm{i} \frac{2\pi n r}{N}}}{N} \frac{1}{2\pi\mathrm{i}}\oint_{\lvert \tilde z\rvert=1} \mathrm{d} \tilde z \, \frac{\tilde z^{(j-\ell)-n-1}}{h(\tilde z)}
          \ketbra{j}{\ell}.
\end{align*}
Using $\frac{1}{N} \sum_{r=1}^N e^{\mathrm{i} \frac{2\pi n r}{N}} = \delta_{n,m N}$ for $m\in\mathbb{Z}$ gives rise to the overall expression
\begin{align}
   \chi_\mathrm{pbc}
      & = \sum_{j,\ell} 
          \sum_{m=-1}^\infty \frac{1}{2\pi\mathrm{i}}\oint_{\lvert \tilde z\rvert=1} \mathrm{d} \tilde z \, \frac{\tilde z^{(j-\ell)-m N-1}}{h(\tilde z)}
          \ketbra{j}{\ell}. \label{eq:chiGenInt}
\end{align}
Here, we have used the fact that since $h(z)$ can at most have $N$ zeros, the sum only starts from $m=-1$.
It follows from Cauchy's principle~\cite{Trefethen2014} that
\begin{align}
   \chi_\mathrm{pbc}
      = & \sum_{j,\ell} [I_{j-\ell} + \varepsilon_{j-\ell}(N)]\ketbra{j}{\ell} \label{eq:MpbcInvTopAndExp}
\end{align}
with
\begin{align}
I_{n}\equiv \sum_{m=-1}^0 \frac{1}{2\pi\mathrm{i}}\oint_{\lvert \tilde z\rvert=1} \mathrm{d} \tilde z \, \frac{\tilde z^{n-m N-1}}{h(\tilde z)} \label{eq:In}
\end{align}
and ${\varepsilon_{n}(N)= \mathcal{O}(c^{-N})}$ an exponentially small correction with some $\lvert c\rvert >1$.
We have obtained exact expressions for $I_n$ and $\varepsilon_n$ with the residue theorem for generating functions of the form~\eqref{eq:genFuncChain}, and we give the results in Methods.
$I_n$ is a function of the zeros of $h(z)$, cf. Eq.~\eqref{eq:zpm}, and thus of the winding number~\eqref{eq:winding}, since the number of zeros within the unit circle determines the contributions to the integral~\eqref{eq:In}, cf. Fig.~\ref{fig:topPhaseDiag}~(c).
This directly connects $\chi_\mathrm{pbc}$ to the winding number.
$I_n$ is at most $\mathcal{O}(1)$ and is illustrated in Fig.~\ref{fig:MobcInverse}, so no significant amplification takes place under PBC.

Moving on to OBC, we express
\begin{align}
M_\mathrm{obc} & = M_\mathrm{pbc} - (\mu_1 \ketbra{1}{N} + \mu_{-1} \ketbra{N}{1})
\end{align}
subtracting the corners of the matrix corresponding to PBC.
To calculate the influence of this change in boundary conditions, we import the following mathematical result~\cite{Miller1981}: The matrix inverse of the sum of an invertible matrix $M$ and a rank-one matrix $E_j$ can be calculated from ${(M+E_j)^{-1} = M^{-1} - \frac{1}{1+g_j}(M^{-1}E_j M^{-1})}$ with $g_j=\mathrm{tr}\,(M^{-1}E_j)$. Applying the formula recursively in two stages, with $E_1=\mu_1\ketbra{1}{N}$ and $E_2=\mu_{-1}\ketbra{N}{1}$, we obtain an analytic expression for $\chi_\mathrm{obc}=M_\mathrm{obc}^{-1}$. Within topologically non-trivial regimes, it simplifies to
\begin{widetext}
\begin{align}
   S(\omega)-\mathbb{1}\propto
   \chi_\mathrm{obc}
      = & 
          \underbrace{
          \sum_{j,\ell=1}^N I_{j-\ell}\ketbra{j}{\ell}
          }_\text{PBC background}
          + 
          \underbrace{\sum_{j,\ell=1}^N
          \left[\frac{\mu_1 I_{j-N} I_{1-\ell}}{1+g_1}
          +
          \frac{\mu_{-1} I_{j-1} I_{N-\ell}}{1+g_2}\right]
          \ketbra{j}{\ell}
          }_\text{directional amplification}
          +
          \underbrace{
          \sum_{j,\ell}\mathcal{O}\left(z_\pm^{\nu N+[\nu(j-\ell)+N]\bmod N}\right)\ketbra{j}{\ell}
          }_\text{exponentially small correction}
          \label{eq:mainResult}
\end{align}
with $g_1 = -\mu_1 (I_{1-N}+\varepsilon_{1-N}(N))$ and $g_2 = -\mu_{-1} (I_{N-1}+\varepsilon_{N-1}(N))$.
\end{widetext}
Eq.~\eqref{eq:mainResult} is one of our central results. The susceptibility $\chi_\mathrm{obc}$ has three contributions: a PBC background equal to $\chi_\mathrm{pbc}$, cf. Eq.~\eqref{eq:MpbcInvTopAndExp}, a term giving rise to directional amplification, and an exponentially small correction. For $N\gg 1$ only the second term dominates due to the division by $(1+g_j)$: for $\nu=+1$ the term $(1+g_1)$ is exponentially small, while ${(1+g_2)}=\mathcal{O}(1)$, and vice versa for $\nu=-1$. This traces back to the values of $\mu_1 I_{1-N}$ and $\mu_{-1}I_{N-1}$ in the definitions of $g_j$, which sensitively depend on $\nu$. One of the $g_j$ is exactly $-1$ if $\nu\neq0$ only leaving $\varepsilon_{1-N}$ or $\varepsilon_{N-1}$ in the denominator.
This exponentially small denominator gives rise to amplification.
We obtain the following expressions for $\nu=+1$ corresponding to ${\lvert z_\pm\rvert<1}$
\begin{align}
   \chi_\mathrm{obc}
   = &
       \sum_{j,\ell=1}^N I_{j-\ell}\ketbra{j}{\ell} -
       \frac{1}{\varepsilon_{1-N}}
       \sum_{j,\ell=1}^N
          I_{j-N} I_{1-\ell}
       \ketbra{j}{\ell} \notag \\
     & +
       \sum_{j,\ell}\mathcal{O}\left(z_-^{j-\ell+N-1}\right)\ketbra{j}{\ell},
       \label{eq:matrixInverseTop}
\end{align}
and for $\nu=-1$ corresponding to $\lvert z_\pm\rvert>1$
\begin{align}
   \chi_\mathrm{obc}
   = &
       \sum_{j,\ell=1}^N I_{j-\ell}\ketbra{j}{\ell} -
       \frac{1}{\varepsilon_{N-1}}
       \sum_{j,\ell=1}^N
          I_{j-1} I_{N-\ell}
       \ketbra{j}{\ell} \notag \\
     & +
       \sum_{j,\ell}\mathcal{O}\left(z_+^{j-\ell-N+1}\right)\ketbra{j}{\ell} \label{eq:matrixInverseTop2}
\end{align}
with
\begin{align}
   \frac{1}{\varepsilon_{\nu(1-N)}}
      & = \nu \mu_1(z_+-z_-) \left[\frac{z_+^{\nu(N+1)}}{1-z_+^{\nu N}} - \frac{z_-^{\nu(N+1)}}{1-z_-^{\nu N}}\right]^{-1}. \label{eq:epsilonExact}
\end{align}
As we show in Fig.~\ref{fig:MobcInverse}~(c) and (d), the above expansions for $\chi_\mathrm{obc}$ converge exponentially fast to the exact result within the whole topologically non-trivial regime, and already yield high accuracy for systems as small as $N=2$ in the vicinity of the EP, where they become exact.
For instance, at $N=2$ for $\theta=\frac{\pi}{2}$, $\mathcal{C}=2.06$, and $\Lambda=2$ the relative error of $\lvert(\chi_\mathrm{obc})_{N,1}\rvert$ is only $3.3\,\%$.
The region of small relative error, Fig.~\ref{fig:MobcInverse}~(d), rapidly extends as $N$ increases, converging faster within the dynamically stable regime and more slowly close to the boundary. \\
Expanding $\varepsilon_{\nu(1-N)}$ of Eq.~\eqref{eq:epsilonExact} for large $N$ and $\lvert z_+\rvert$ sufficiently different from $\lvert z_-\rvert$, we obtain
\begin{align}
   \frac{1}{\varepsilon_{\nu(N-1)}} & \cong
     \mu_1(z_+-z_-) \,  z_\pm^{-\nu(N+1)} \label{eq:ampl}
\end{align}
in which we choose $z_+$ in the expansion for $\nu=+1$ and $z_-$ for $\nu=-1$.

The susceptibility $\chi_\mathrm{obc}$ determines the behavior of $S(\omega)$ according to Eq.~(\ref{eq:scatMat}).
We identify $1/\varepsilon_{\nu(1-N)}$ as the contribution giving rise to amplification, as it is directly related to the gain~\eqref{eq:gain}, which asymptotically grows exponentially with the system size
\begin{align}
   \mathcal{G}_{\nu=\pm1} \cong \, \gamma^2 \, \frac{\lvert \mu_0^2-4\mu_1\mu_{-1}\rvert}{\lvert\mu_{\pm 1}\rvert^4} \, \lvert z_\pm\rvert^{-2\nu (N+1)}, \label{eq:gainZpm}
\end{align}
and at the EP, $\mathcal{G}_{\nu=\pm1}\cong\frac{\gamma^2}{\lvert\mu_{\pm1}\rvert^2}\left\lvert\frac{\mu_{\pm1}}{\mu_0}\right\rvert^{2N}$.
In the thermodynamic limit, $N\to\infty$, $\mathcal{G}$ diverges within non-trivial regimes, but stays finite in trivial regimes.
We can also give the asymptotic expression for the reverse gain. The leading order contribution stems from the PBC background, $I_{\nu(N-1)}$, and therefore $\bar{\mathcal{G}}$ decreases exponentially with $N$
\begin{align}
    \bar{\mathcal{G}}_{\nu=\pm1}
    \cong \gamma^2 \, \frac{1}{\lvert \mu_0^2-4\mu_1\mu_{-1}\rvert^2} \lvert z_{\mp}\rvert^{2\nu(N+1)}, \label{eq:revGainZpm}
\end{align}
and at the EP, $\bar{\mathcal{G}}=0$ exactly.
These expressions also converge exponentially fast and are most practical starting from $N\approx5$.

In general, the individual elements of $\chi_\mathrm{obc}$ and therefore the scattering matrix~\eqref{eq:scatMat} are formed by the terms $I_{j-N} I_{1-\ell}$, and $I_{j-1} I_{N-\ell}$, according to Eqs.~\eqref{eq:matrixInverseTop} and \eqref{eq:matrixInverseTop2}, respectively, which give rise to directionality. Since $I_n$ decreases approximately exponentially with $n$ and is defined modulo $N$, the products of the different $I_n$ only leave one matrix element that contributes significantly, see Fig.~\ref{fig:MobcInverse}. This is the one determining the gain~\eqref{eq:gainZpm}.

In trivial topological regimes we obtain more cumbersome combinations of $I_n$ and $1/(1+g_j)$, but $(1+g_j)=\mathcal{O}(1)$, so no amplification takes place. However, as we can see from the scattering matrices displayed in Fig.~\ref{fig:topPhaseDiag}~(b), directionality is still possible.

\subsection*{Applications --- design of multimode directional amplifiers}
\label{sec:applications}

So far, we have focused on the driven-dissipative chain~\eqref{eq:dynEq}, however, the results of Eqs.~\eqref{eq:mainResult} to \eqref{eq:revGainZpm} apply more generally to systems with nearest-neighbor couplings. We can map any system with a generating function of the form~\eqref{eq:genFunction} to the parameters of the driven-dissipative chain, i.e.~$\mathcal{C}$, $\Lambda$, $\tilde\omega$ and $\theta$, and apply all of our previous results.
However, the physical interactions giving rise to amplification and indeed the amplified observables may be very different from those of the driven-dissipative chain.
We illustrate this by applying our topological framework to several models for phase preserving and phase sensitive amplifiers.
Remarkably, the expressions for the scattering matrix~\eqref{eq:mainResult}, the `phase diagram' Fig.~\ref{fig:topPhaseDiag}~(b), the gain~\eqref{eq:gainZpm} and the reverse gain~\eqref{eq:revGainZpm} in Fig.~\ref{fig:topPhaseDiag}~(a) apply \textit{mutatis mutandis}. Hence, we obtain the same exponential growth and attenuation with $N$ for gain and reverse gain, respectively, without any explicit calculations.

\begin{figure}[t]
\centering
\includegraphics[width=.48\textwidth]{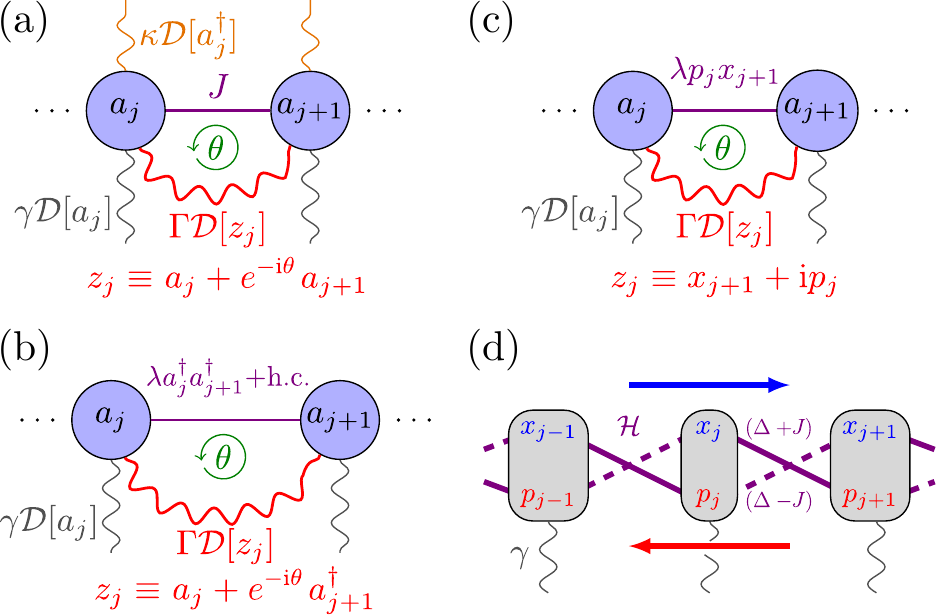}
\caption{
\textbf{Some directional amplifiers covered by our framework.}
(a)~Our driven-dissipative cavity chain~Eqs.~\eqref{eq:dynEq}, (b)~a phase insensitive amplifier with parametric interactions, (c)~a phase sensitive amplifier amplifying both $x$ and $p$ quadratures separately, and (d) a bosonic Kitaev chain~\cite{McDonald2018}, which amplifies $x$ and $p$ quadratures in opposite directions.
(b) and (c) are based on~\cite{Metelmann2015, Metelmann2017}.
All of these amplifiers can be analyzed with our topological framework.
}
\label{fig:exampleAmplifiers}
\end{figure}

First, we focus on the phase preserving amplifier proposed by Metelmann and Clerk~\cite{Metelmann2015, Metelmann2017} and sketched in Fig.~\ref{fig:exampleAmplifiers}~(b). We consider the generalization of their two-mode proposal to a chain of $N$ cavities. Two neighboring modes $a_j$ and $a_{j+1}$ are coupled both via the coherent parametric interaction $\lambda a^\dagger_j a^\dagger_{j+1}+\lambda^* a_j a_{j+1}$ and through the non-local dissipator $\mathcal{D}[ a_j+e^{-\mathrm{i}\theta} a_{j+1}^\dagger]$.
Gauge freedom allows us to absorb the phase into $\lambda$; however, we focus on the case of imaginary $\lambda$, i.e.~$\lambda=\mathrm{i}\lvert\lambda\rvert$, which ensures that the amplifier does not couple different quadratures and therefore is phase insensitive.
The equations of motion for the field quadratures $x_j\equiv(a_j+a_j^\dagger)/\sqrt{2}$ and $p_j\equiv-\mathrm{i}(a_j-a_j^\dagger)/\sqrt{2}$ are then given by
\begin{align}
   \langle \dot x_\ell \rangle = -\frac{\gamma}{2} \langle x_\ell\rangle & + \left(\lvert\lambda\rvert-\frac{\Gamma}{2}\right) \langle x_{\ell+1}\rangle \notag\\
   & + \left(\lvert\lambda\rvert+\frac{\Gamma}{2}\right)\langle x_{\ell-1}\rangle - \sqrt{\gamma} \langle x_{\ell,\mathrm{in}}\rangle \label{eq:eomPhaseInsX} \\
   \langle \dot p_\ell \rangle = -\frac{\gamma}{2} \langle p_\ell\rangle & - \left(\lvert\lambda\rvert-\frac{\Gamma}{2}\right) \langle p_{\ell+1}\rangle \notag\\
   & - \left(\lvert\lambda\rvert+\frac{\Gamma}{2}\right)\langle p_{\ell-1}\rangle - \sqrt{\gamma} \langle p_{\ell,\mathrm{in}}\rangle.
   \label{eq:eomPhaseInsP}
\end{align}
From the equations above we can directly read off the generating function for the two quadratures. Introducing $\mathcal{C}\equiv 4\lvert\lambda\rvert/\gamma$ and $\Lambda\equiv 2\Gamma/\gamma$, we find
\begin{align}
   h_x(k) & \propto -1 + \mathcal{C} \cos k + \mathrm{i} \Lambda \sin k \\
   h_p(k) & \propto -1 - \mathcal{C} \cos k - \mathrm{i} \Lambda \sin k, \label{eq:hP}
\end{align}
Notice that $x$ and $p$ quadratures have the same generating function up to the sign of the oscillating terms, which reflects the phase conjugating property of the amplifier: $x$ and $p$ quadratures are amplified with the same gain, but the $p$ quadrature exits with a $\pi$ phase shift, i.e.~a negative sign, at the output. Nevertheless, the amplifier is still considered to be phase insensitive according to \cite{Caves1982}. The minus sign has no impact on the topological regimes, since $\cos k$ in Eq.~\eqref{eq:hP} takes both positive and negative values as $h_p$ winds around the origin, and we obtain the same regimes for $x$ and $p$ quadratures according to Eq.~\eqref{eq:topCondNoDet}:
$\nu=0$ for $\mathcal{C}<1$,
$\nu=+1$ for $\mathcal{C}>1$ and $\Lambda>0$.
We have set $\theta=\frac{\pi}{2}$ for $\nu=-1$ and $\theta=\frac{3\pi}{2}$ for $\nu=+1$ in Eq.~\eqref{eq:topCondNoDet}, since $\theta$ is defined as the phase difference between real and imaginary part.

Therefore, gain and reverse gain for the quadratures of the phase insensitive amplifier, Eq.~\eqref{eq:eomPhaseInsX} and \eqref{eq:eomPhaseInsP}, are given by Fig.~\ref{fig:topPhaseDiag}~(a) with $\mathcal{C}=4\lvert\lambda\rvert/\gamma$.  Furthermore, the scattering matrices $S_x(\omega)$ and $S_p(\omega)$ linking ${\bf x_\mathrm{out}}=S_x{\bf x_\mathrm{in}}$ and ${\bf p_\mathrm{out}}=S_x{\bf p_\mathrm{in}}$ with
${\bf x_\mathrm{in/out}}\equiv(\langle x_{1,\mathrm{in/out}}\rangle,\dots,\langle x_{N,\mathrm{in/out}}\rangle)^\mathrm{T}$ and ${\bf p_\mathrm{in/out}}\equiv(\langle p_{1,\mathrm{in/out}}\rangle,\dots,\langle p_{N,\mathrm{in/out}}\rangle)^\mathrm{T}$
are given by Eq.~\eqref{eq:mainResult}.
Since the generating functions are the same up to the sign conjugation, $\lvert S_x(0)\rvert^2=\lvert S_p(0)\rvert^2$; off resonance, analogous considerations lead to $\lvert S_x(\omega)\rvert^2=\lvert S_p(\omega)\rvert^2$. Beyond that, the scattering matrices $\lvert S_x(0)\rvert^2$, $\lvert S_p(0)\rvert^2$ take the same form as the insets in Fig.~\ref{fig:topPhaseDiag}~(b) with $\theta=\frac{\pi}{2}$ and $\theta=\frac{3\pi}{2}$.
Furthermore, the asymptotic scaling of the gain is given by $\mathcal{G}_{\nu=\pm1}\propto \lvert z_\mp\rvert^{-2\nu N}$ and of the reverse gain by $\bar{\mathcal{G}}_{\nu=\pm1}\propto \lvert z_\pm\rvert^{2\nu N}$ according to Eqs.~\eqref{eq:gainZpm} and \eqref{eq:revGainZpm}, respectively.
This demonstrates the power of the framework: we can determine the properties of a physically very different amplifier consisting now generally of $N$ modes without numerically calculating the scattering matrix.

Next, we examine the phase sensitive amplifier proposed in~\cite{Metelmann2015,Metelmann2017}. It couples the field quadratures  via the coherent interaction $\lambda p_jx_{j+1}$ and the dissipator $\Gamma\mathcal{D}[x_{j+1}+\mathrm{i} p_j]$.
We again consider the generalization to a chain of $N$ modes and obtain the equations of motion
\begin{align}
   \langle\dot x_\ell\rangle & = -\frac{\gamma}{2} \langle x_\ell\rangle - (\Gamma - \lambda) \langle x_{\ell+1}\rangle - \sqrt{\gamma} \langle x_{\ell,\mathrm{in}}\rangle \label{eq:eomPhaseSensX}\\
   \langle\dot p_\ell\rangle & = -\frac{\gamma}{2} \langle p_\ell\rangle - (\Gamma + \lambda) \langle p_{\ell-1}\rangle - \sqrt{\gamma} \langle p_{\ell,\mathrm{in}}\rangle. \label{eq:eomPhaseSensP}
\end{align}
The equations for $x$ and $p$ quadratures decouple and therefore, we consider them separately.

\begin{figure}[t]
   \centering
   \includegraphics[width=.48\textwidth]{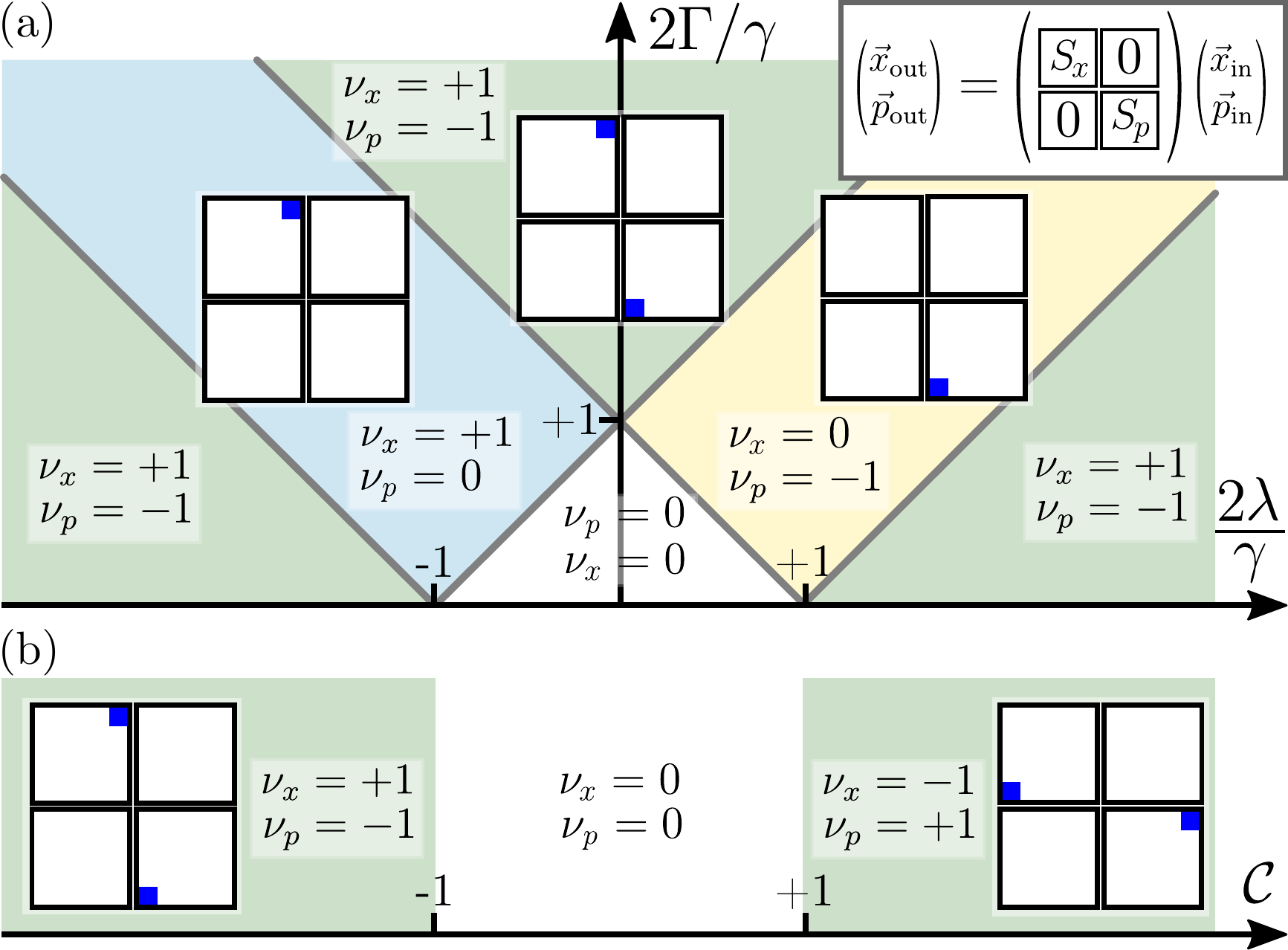}
   \caption{\textbf{Topological `phase diagram' for the phase sensitive amplifier, Eqs.~\eqref{eq:eomPhaseSensX} and \eqref{eq:eomPhaseSensP}, and the bosonic Kitaev chain, Eq.~\eqref{eq:KitaevHam}.}
   Topological regimes for (a) the phase sensitive amplifier of Fig.~\ref{fig:exampleAmplifiers}~(c), and (b) the bosonic Kitaev chain of Fig.~\ref{fig:exampleAmplifiers}~(d).
   We sketch the scattering matrix $\lvert S(0)\rvert^2$, whereby the blue rectangles indicate the dominant matrix elements.
   It is block diagonal for both systems (see inset), since their equations of motion decouple, with blocks $S_x$ and $S_p$ addressing $x$ and $p$ quadratures, respectively.
   Each block is obtained from Eq.~\eqref{eq:mainResult}.
   In (a), depending on the parameters $\Gamma$ and $\lambda$, either the $x$ quadratures are in a non-trivial regime, the $p$ quadratures, both or none. The central white region is topologically trivial, so the gain is $\mathcal{O}(1)$.
   In (b), either both $x$ and $p$ quadratures are in a non-trivial regime with $\nu_x=-\nu_p$, or $\nu_x=\nu_p=0$.
   Since the winding numbers $\nu_x$ and $\nu_p$ have opposite sign,
      the amplification direction is the reverse for the two quadratures.
   }
\label{fig:phaseSensDiag}
\end{figure}

Defining $\mathcal{C}_\pm\equiv2(\Gamma\pm\lambda)/\gamma$ with the positive sign for $p$ and the negative sign for $x$, the generating functions take the form
\begin{align}
   h_x(k) & \propto -1 - \mathcal{C}_- \cos k - \mathrm{i} \mathcal{C}_-\sin k \\
   h_p(k) & \propto -1 - \mathcal{C}_+ \cos k + \mathrm{i} \mathcal{C}_+ \sin k.
\end{align}
We obtain the following topological regimes from condition~\eqref{eq:topCondNoDet} with $\theta=\frac{3\pi}{2}$:
$\nu_x=+1$ for $\lvert\mathcal{C}_-\rvert>1$,
$\nu_x=0$ for $\lvert\mathcal{C}_-\rvert<1$;
and with $\theta=\frac{\pi}{2}$:
$\nu_p=-1$ for $\lvert\mathcal{C}_+\rvert>1$,
$\nu_p=0$ for $\lvert\mathcal{C}_+\rvert<1$,
where $\nu_x$ and $\nu_p$ refer to the winding numbers for $x$ and $p$ quadratures, respectively.
As we illustrate in Fig.~\ref{fig:phaseSensDiag}~(a), depending on the regime, both quadratures, only one of them, or none, are amplified. The amplification direction for $x$ and $p$ quadratures is the reverse. We again calculate the scattering matrices $S_x$ and $S_p$ for $x$ and $p$ from Eq.~\eqref{eq:mainResult} and show some as insets in Fig.~\ref{fig:phaseSensDiag}~(a). Analogously, the gain and the reverse gain are obtained from Eqs.~\eqref{eq:gainZpm} and \eqref{eq:revGainZpm}, respectively. The gain follows the same behavior as Fig.~\ref{fig:topPhaseDiag}~(a).

Finally, we consider the `bosonic Kitaev chain' proposed in Ref.~\cite{McDonald2018} and illustrated in Fig.~\ref{fig:exampleAmplifiers}~(d), for which $x$ and $p$ quadratures are amplified in opposite directions. This also follows straightforwardly from our topological framework.
The Hamiltonian
\begin{align}
   \mathcal{H}
   & = \frac{1}{2} \sum_j [(\Delta-J) x_{j+1}p_j+(\Delta+J)p_{j+1}x_j] \label{eq:KitaevHam}
\end{align}
together with on-site dissipator $\gamma\mathcal{D}[a_j]$
gives rise to the following equations of motion for the system's quadratures
\begin{align}
   \langle\dot x_\ell\rangle & = -\frac{\gamma}{2}\langle x_{\ell}\rangle + \frac{J+\Delta}{2} \langle x_{\ell-1}\rangle - \frac{J-\Delta}{2}\langle x_{\ell+1}\rangle -\sqrt{\gamma}\langle x_{\ell,\mathrm{in}}\rangle \notag \\
   & \equiv \sum_j H_{\ell,j} \langle x_j \rangle -\sqrt{\gamma}\langle x_{\ell,\mathrm{in}}\rangle \label{eq:eomXKitaev} \\
   \langle\dot p_\ell\rangle & = -\frac{\gamma}{2}\langle p_{\ell}\rangle + \frac{J-\Delta}{2} \langle p_{\ell-1}\rangle - \frac{J+\Delta}{2} \langle p_{\ell+1}\rangle -\sqrt{\gamma}\langle p_{\ell,\mathrm{in}}\rangle \notag \\
   & \equiv \sum_j (-H^\mathrm{T})_{\ell,j} \langle p_j \rangle -\sqrt{\gamma}\langle p_{\ell,\mathrm{in}}\rangle. \label{eq:eomPKitaev}
\end{align}
We have added coherent driving to obtain the input terms in Eqs.~\eqref{eq:eomXKitaev} and \eqref{eq:eomPKitaev} and cast them into the same form as Eqs.~\eqref{eq:dynEq}.

As we can see from the last lines of Eqs.~\eqref{eq:eomXKitaev} and \eqref{eq:eomPKitaev}, the dynamic matrix governing the evolution of the $p$ quadratures is the negative transpose of that of the $x$ quadratures. On the level of the generating functions, this translates into a change in the sign of the winding number within topologically non-trivial regimes.
Defining $\mathcal{C}\equiv2\Delta/\gamma$ and $\Lambda\equiv2J/\gamma$, the generating functions are
\begin{align}
   h_x(k) & \propto -1 + \mathcal{C} \cos k - \mathrm{i} \Lambda \sin k \label{eq:genFuncKitaevX} \\
   h_p(k) & \propto -1 - \mathcal{C} \cos k - \mathrm{i} \Lambda \sin k \label{eq:genFuncKitaevP}.
\end{align}
Assuming $\Lambda>0$, we obtain from condition~\eqref{eq:topCondNoDet}:
$\nu_x=0$ and $\nu_p=0$ for $\lvert\mathcal{C}\rvert<1$,
$\nu_x=-1$ and $\nu_p=+1$ for $\mathcal{C}>1$,
$\nu_x=+1$ and $\nu_p=-1$ for $\mathcal{C}<-1$, cf. Fig.~\ref{fig:phaseSensDiag}~(b).
The non-trivial cases correspond to setting $\theta=\frac{\pi}{2}$ for $x$ and $\theta=\frac{3\pi}{2}$ for $p$ quadratures for $\mathcal{C}>1$, or vice versa for $\mathcal{C}<-1$, in the `phase diagram' Fig.~\ref{fig:topPhaseDiag}~(b) and the gain Fig.~\ref{fig:topPhaseDiag}~(a).
As for the previous examples, we obtain the scattering matrices from Eq.~\eqref{eq:mainResult} and illustrate them in Fig.~\ref{fig:phaseSensDiag}~(b).
Since the winding numbers for $x$ and $p$ quadratures have opposite sign they are amplified in reverse directions.
Gain and reverse gain follow from Eqs.~\eqref{eq:gainZpm} and \eqref{eq:revGainZpm}, respectively.

\section*{Discussion}

In this work we have developed a framework based on the topology of the dynamic matrix to predict and describe directional amplification in driven-dissipative systems.
In contrast to topological states of matter for closed systems, we have introduced the winding number~\eqref{eq:winding} as topological invariant based on the spectrum of the dynamic matrix --- the generating function~\eqref{eq:genFunction}.
We have shown that non-trivial values of the winding number have a directly observable consequence expressed in the scattering matrix~\eqref{eq:scatMatIntroduction}, and we have established a one-to-one correspondence between non-trivial topology and directional amplification.
One of our main results is the `phase diagram' for the scattering matrix, Fig.~\ref{fig:topPhaseDiag}~(b), that associates topologically non-trivial parameter regimes with directional amplification.
We have obtained an analytic expression for the scattering matrix~(\ref{eq:scatMat}) in Eq.~(\ref{eq:mainResult}), the gain~(\ref{eq:gainZpm}) and the reverse gain~\eqref{eq:revGainZpm} in the case of nearest-neighbor couplings and have revealed an exponential scaling of the gain with the number of sites within topologically non-trivial phases, while the reverse gain is exponentially suppressed. In the limit of an infinite chain, completely directional amplification is obtained within the whole topological regime.
Our result for the scattering matrix~\eqref{eq:mainResult} already yields high accuracy for systems as small as $N=2$ in the vicinity of the EP, where it is exact, and it converges exponentially fast within the whole topologically non-trivial regime.
Therefore, directional amplification can be seen as a proxy of non-trivial topology, formally defined only in the limit $N\to\infty$, even in very small systems, which makes our work relevant for state-of-the art devices such as~\cite{Mercier2019}.
Furthermore, we have demonstrated the generality of our results and shown how four systems each with different coherent and dissipative interactions can be analyzed with our topological framework.
One of our key assumption is translational invariance. However, we still expect our results to serve as good approximation when the terms breaking translational invariance are sufficiently small.
Another way to go beyond our assumptions is, for instance, to add parametric interactions to Eq.~\eqref{eq:dynEq}. This yields two rather than one complex band, and we have to modify our main result~\eqref{eq:mainResult}.
Interactions beyond nearest neighbors yield yet another form of the dynamic matrix which leads to higher winding numbers. This necessaitates additional terms in our decomposition of the scattering matrix, Eq.~\eqref{eq:mainResult}. These extensions will be addressed in future work.

Suitable platforms for implementation include superconducting circuits~\cite{Bergeal2010, Abdo2013DirAmp}, optomechanics~\cite{Aspelmeyer2014}, photonic crystals~\cite{Ozawa2019} and nanocavity arrays~\cite{Rider2019}, as well as topolectric circuits~\cite{Lee2018, Kotwal2019} and mechanical meta-materials~\cite{Nash2015, Huber2016, Ghatak2019Mech}.
On a fundamental level, our analysis sheds light on the role of topology in open quantum systems~\cite{Bardyn2013} and is of direct relevance for the study of non-Hermitian topology~\cite{MartinezAlvarez2018, Gong2018, Ghatak2019, Kawabata2018}, where our framework predicts immediate physical and observable consequences for a topological invariant.

\section*{Methods}

\subsection*{Exact expressions}
\label{sec:residues}

We give here the exact expressions for $I_n$ and $\epsilon_n$ arising in the derivation of our main results Eqs.~\eqref{eq:mainResult} to \eqref{eq:revGainZpm} --- the one-to-one correspondence between a non-trivial winding number and directional amplification.
$\chi_\mathrm{obc}$ is crucially determined by
$I_{n}\equiv \sum_{m=-1}^0 \frac{1}{2\pi\mathrm{i}}\oint_{\lvert \tilde z\rvert=1} \mathrm{d} \tilde z \, \frac{\tilde z^{n-m N-1}}{h(\tilde z)}$,
see Eq.~\eqref{eq:mainResult}. We can calculate it exactly for generating functions~\eqref{eq:genFunction} using the residue theorem. For that purpose, we use the general Leibniz rule and find the residues with $r(n) \equiv (\nu n+N) \bmod N$
\begin{itemize}
\item $\nu\neq 0$, i.e.~either $\lvert z_\pm\rvert>1$ or $\lvert z_\pm\rvert<1$, and $z_+\neq z_-$
\begin{align}
 I_n & = \frac{\nu}{\mu_1} \frac{z_+^{\nu \lvert r(n)\rvert }-z_-^{\nu \lvert r(n)\rvert}}{z_+ - z_-}
\end{align}
\item $\nu\neq 0$ and $z_+= z_-$
\begin{align}
 I_n & =
    \begin{cases}
       \frac{1}{\mu_1} \lvert r(n)\rvert z_+^{\nu\lvert r(n)\rvert-1}: & n\neq 0 \\
       0 & n=0
    \end{cases}
\end{align}
\item
    $\nu=0$: $\lvert z_+\rvert < 1$ and $\lvert z_-\rvert > 1$ or $\lvert z_+\rvert > 1$ and ${\lvert z_-\rvert < 1}$
\begin{align}
    I_n & =
       \begin{cases}
          \pm \frac{1}{\mu_1}\frac{z_\pm^{\lvert n\rvert}}{z_+-z_-} & : n \geq 0 \\
          \pm \frac{1}{\mu_1}\frac{z_\mp^{-\lvert n\rvert}}{z_+-z_-} & : n < 0.
       \end{cases}
\end{align}
\end{itemize}
One important feature of this expression within topological phases is $I_0=0$. This allows us to simplify $\chi_\mathrm{obc}$ to yield Eq.~\eqref{eq:mainResult}.

We also employ the residue theorem to calculate the correction $\varepsilon_n(N)$ exactly rewriting the sum as geometric series and inserting the calculated residues
\begin{align}
   \varepsilon_{n}(N)
      & \equiv \sum_{m=1}^\infty \frac{1}{2\pi\mathrm{i}}\oint_{\lvert \tilde z\rvert=1} \mathrm{d} \tilde z \, \frac{\tilde z^{n-m N-1}}{h(\tilde z)} \notag \\
      & = \frac{\nu}{\mu_1(z_+-z_-)} \left(\frac{z_+^{\nu(N+r(n))}}{1-z_+^{\nu N}} - \frac{z_-^{\nu(N+r(n))}}{1-z_-^{\nu N}}\right) \notag \\
   \frac{1}{\varepsilon_{\nu(1-N)}}
      & \cong \mu_1(z_+-z_-) z_\pm^{-\nu(N+1)},
\end{align}
in which $\pm$ is chosen according to the winding number: $z_+$ for $\nu=+1$ and $z_-$ for $\nu=-1$.

\begin{figure}[t]
   \centering
   \includegraphics[width=.48\textwidth]{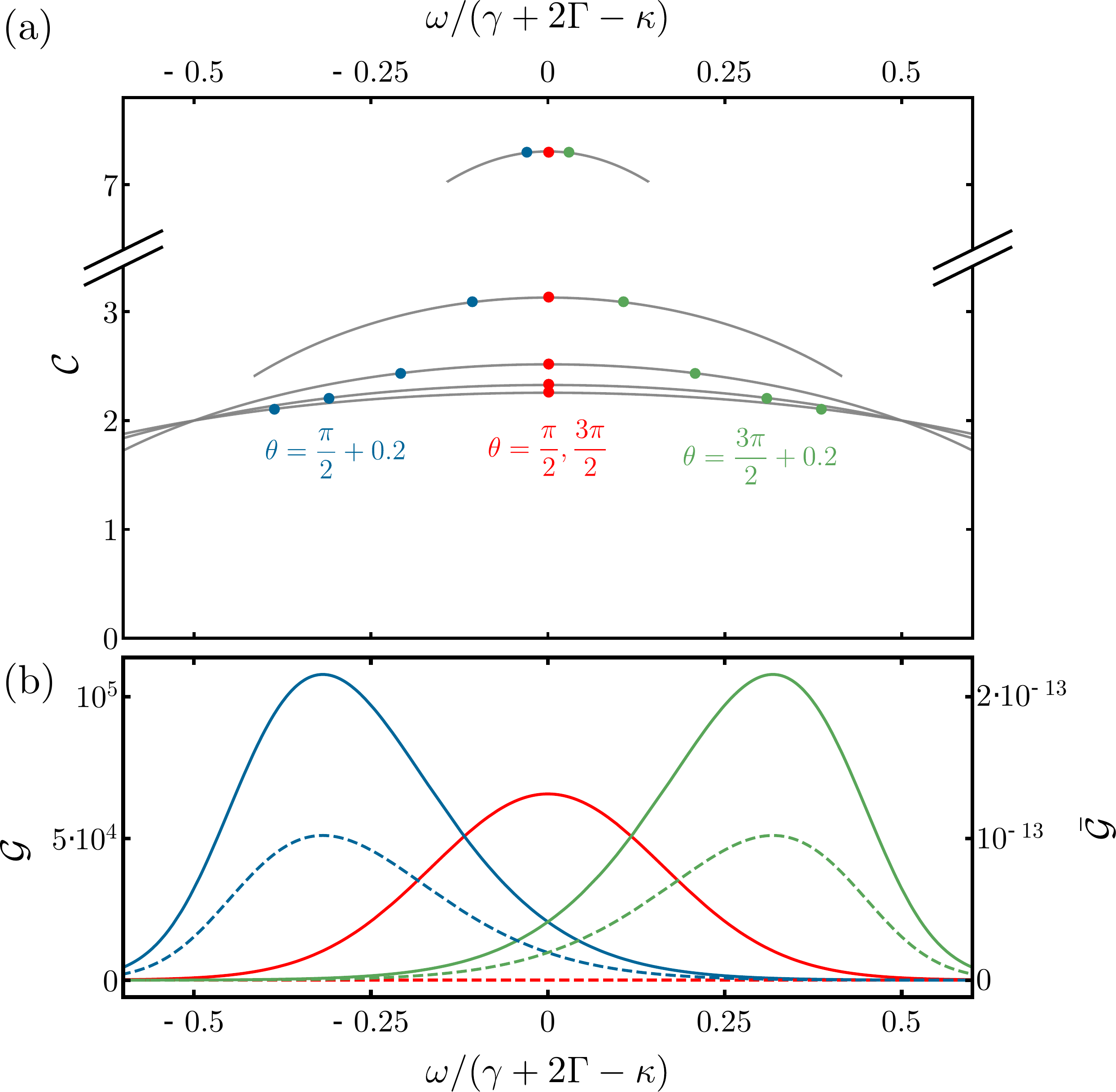}
   \caption{
   \textbf{Gain and reverse gain}.
   In (b)~we show the gain $\mathcal{G}(\omega)$ of Eq.~\eqref{eq:gainOmega} (solid line) and reverse gain $\bar{\mathcal{G}}(\omega)$ of Eq.~\eqref{eq:revGainOmega} (dashed line) for $\mathcal{C}=1.5$ for different $\theta$.
   The gain is a product of Lorentzians of width $(\gamma+2\Gamma-\kappa)$, cf. Eq.~\eqref{eq:LorentzGain}.
   The position of the divergences at larger $\mathcal{C}$ in (a) sets the position of the peak in gain and reverse gain.
   For $\theta=\frac{\pi}{2},\frac{3\pi}{2}$ the peak is centered around $\omega=0$.
   }
   \label{fig:gainAndPoles}
\end{figure}

\subsection*{Determining the exceptional point (EP)}
The value of the EP can be extracted analytically for all $N$. At the EP, eigenvalues and eigenvectors coalesce. The dynamic matrix, Eq.~\eqref{eq:dynEq}, is a Toeplitz matrix, for which there exists an analytic expression for both eigenvalues and eigenvectors~\cite{Willms2008}, see Eq.~\eqref{eq:eigenvals}. From this expression it is clear, that the eigenvalues can only coalesce when either
$\mathrm{i} J=-\frac{e^{\mathrm{i}\theta}\Gamma}{2}$
or
$\mathrm{i} J=-\frac{e^{-\mathrm{i}\theta}\Gamma}{2}$,
in which case the dynamic matrix becomes an upper (lower) triangular matrix with only the diagonal and super-(sub-)diagonal non-zero. Since all the entries on the respective diagonal and super-(sub-)diagonal are the same, the matrix has rank $1$ and these are indeed exceptional points. We obtain the $N$-fold degenerate right eigenvectors from Gaussian elimination to be either $(1,0,\dots,0,0)^\mathrm{T}$ in the former case or $(0,0,\dots,0,1)^\mathrm{T}$ in the latter case.

\subsection*{Stability and bandwidth of the driven-dissipative cavity chain}
\label{sec:eigvalsSpectral}

Here, we discuss the stability of the driven-dissipative chain as well as the gain $\mathcal{G}(\omega)$ as a function of $\omega$.
Stability requires the real part of all eigenvalues $\lambda_m$ of the dynamic matrix $M_\mathrm{obc}$ to be negative.
The analytic expression for $\lambda_m$ is given by~\cite{Willms2008}
\begin{align}
   \lambda_m
      =
          & \frac{\lvert 2\Gamma+\gamma-\kappa\rvert}{2}
          \left[\vphantom{\left[\frac{m\pi}{N+1}\right]}
             -1+\mathrm{i} \tilde\omega \right. \notag \\
          & + \left.
             \sqrt{\mathcal{C}^2-\Lambda^2+2\mathrm{i}\mathcal{C}\Lambda\cos\theta}
             \cos \left(\frac{m\pi}{N+1}\right)
          \right] \label{eq:eigenvals}
\end{align}
for $\kappa<2\Gamma+\gamma$ and $m=1,\dots,N$.
Larger values of $\Lambda$ extend the stable regime to larger $\mathcal{C}$. In order to obtain a regime which is both stable and amplifying, we require $\Lambda>1$.

The eigenvalues also determine the bandwidth of the gain $\mathcal{G}(\omega)$.
We can write the exact expression for $\mathcal{G}(\omega)$ using~\cite{daFonseca2001}.
Together with Eq.~(\ref{eq:scatMat}) and denoting $\mu_0(\omega)=-1+\mathrm{i} \tilde\omega$ and $\mu_{\pm1}=-\mathrm{i}\Lambda-\mathcal{C}e^{\mp\mathrm{i}\theta}$, we write the gain
\begin{align}
   \mathcal{G}_{\nu=\pm1}(\omega) & = \left\lvert\frac{\mu_{\mp 1}^{N-1}}{(\mu_1\mu_{-1})^{N/2}}\frac{1}{U_N\left(\frac{\mu_0(\omega)}{2\sqrt{\mu_1\mu_{-1}}}\right)}\right\rvert^2 \label{eq:gainOmega}
   \intertext{and the reverse gain}
   \bar{\mathcal{G}}_{\nu=\pm1}(\omega) & = \left\lvert\frac{\mu_{\pm 1}^{N-1}}{(\mu_1\mu_{-1})^{N/2}}\frac{1}{U_N\left(\frac{\mu_0(\omega)}{2\sqrt{\mu_1\mu_{-1}}}\right)}\right\rvert^2, \label{eq:revGainOmega}
\end{align}
in which $U_N$ denotes the Chebyshev polynomial of the second kind.
This expression diverges at the zeros of the Chebyshev polynomials, see peaks in Fig.~\ref{fig:topPhaseDiag}~(b), which satisfy~\cite{AbramowitzStegun1964}
\begin{align}
   \frac{\mu_0(\omega)}{2\sqrt{\mu_1\mu_{-1}}} = \cos\left(\frac{m+1}{N}\pi\right), \label{eq:resCondition}
\end{align}
with $m=1,2,\dots,N$. This is equivalent to $\lambda_m=0$ for at least one eigenvalue, cf.~\eqref{eq:eigenvals}. In principle, this equation has $N$ solutions, however, since $\Re\mu_0(\omega)=-1$, the above condition cannot be fulfilled for all parameters $\Lambda$ and $\theta$, and we only obtain $\lfloor(N/2)\rfloor$ zeros, see Fig.~\ref{fig:gainAndPoles}. A factorization in terms of these zeros lets us write $\mathcal{G}(\omega)$ as product of Lorentzians
\begin{align}
   \mathcal{G}(\omega) \propto \prod_{j=1}^{\lfloor(N/2)\rfloor}\frac{\left(\frac{\gamma+2\Gamma-\kappa}{2}\right)^2}{(\omega-\omega_j)^2+\left(\frac{\gamma+2\Gamma-\kappa}{2}\right)^2}, \label{eq:LorentzGain}
\end{align}
in which the $\omega_j$ can be determined from Eq.~(\ref{eq:resCondition}).
All Lorentzians have the same width set by the effective on-site dissipation $(\gamma+2\Gamma-\kappa)$.
However, if the Lorentzians are centered around distinct $\omega_j$, which is the case if $\theta\neq\frac{\pi}{2}$ and $\theta\neq\frac{3\pi}{2}$, the peak is broadened, see Fig.~\ref{fig:gainAndPoles}.
Therefore, the amplifier has no conventional gain-bandwidth product, which will be the subject of future research.
The reverse gain has the same line shape, but is suppressed by many orders of magnitude --- it is attenuated exponentially with $N$, see Eq.~\eqref{eq:revGainZpm}.

\section*{Data availability}
No datasets were generated or analyzed during the current study.

\section*{Author contributions}
A.N. initiated and directed the project. C.C.W. derived the analytical results with input from M.B. All authors contributed to the writing of the manuscript.

\section*{Competing financial interests} The authors declare no competing interests.

\begin{acknowledgments}
We would like to thank Katarzyna Macieszczak and Daniel Malz for insightful discussions. C.C.W. acknowledges the funding received from the Winton Programme for the Physics of Sustainability and the EPSRC (Project Reference EP/R513180/1).
A.N. holds a University Research Fellowship from the Royal Society and acknowledges additional support from the Winton Programme for the Physics of Sustainability.
We are grateful for the funding received from the European Union's Horizon 2020 research and innovation programme under Grant No. 732894 (FET Proactive HOT).
\end{acknowledgments}



\end{document}